\documentclass[preprintnumbers,amsmath,amssymb,floatfix,11pt,prd,onecolumn,
superscriptaddress,nofootinbib]{revtex4}
\usepackage{latexsym}
\usepackage{epsfig}
\usepackage{amsmath}
\usepackage{epstopdf}
\usepackage{amssymb}
\usepackage{graphicx}
\usepackage{hyperref}
\usepackage{tabularx}
\usepackage{varioref}
\usepackage{placeins}
\newcolumntype{C}[1]{>{\centering\arraybackslash}p{#1}}

\begin{document}
\title{\bf Center of Mass Energy of Three General Geodesic Colliding Particles Around a Kerr-MOG Black Hole}
\author{Ayesha Zakria}\email{ayesha.zakria@bcoew.edu.pk}
\affiliation{Department of Mathematics, Bilquis Post Graduate College for Women, Air University, Islamabad, Pakistan}
\begin{abstract}
{\bf Abstract:} In this research paper, we scrutinize the center of mass energy of the collision for three neutral particles with different rest masses falling freely from rest at infinity in the vicinity of a Kerr-Modified-Gravity black hole. In addition, we deliberate the center of mass energy adjacent to the horizon(s) of an extremal and non-extremal Kerr-Modified-Gravity black hole and demonstrate that a swiftly huge center of mass energy is attainable concealed by few constraints. 
\end{abstract}

 \maketitle
 \newpage
\section{Introduction}
Rotating black holes can accelerate particles to a swiftly huge energy if the angular momentum of any of the particles is fine-tuned to some critical value. An arbitrary high center of mass energy (CME) of the collision for the two particles is subjected to several physical effects. The high energy collision of particles in the vicinity of the horizon(s) can produce high energy and/or superheavy particles. Ba\~{n}ados, Silk and West (BSW) \cite{1} analysed the collision for two particles in the vicinity of a Kerr black hole and settled the CME in the equatorial plane. Grib and Pavlov \cite{2,3,4} demonstrated that very big values of the scattering energy of particles in the center of mass frame can be achieved for an extremal and non-extremal Kerr
black hole. In \cite{5, 6}, the authors mentioned that the arbitrarily huge CME might not be attainable in nature due to the astrophysical limitations i.e., the maximal spin and gravitational radiation.
Lake \cite{7} in addition illuminated that the CME for two colliding particles is divergent at the inner horizon of a non-extremal Kerr black hole. The collision in the innermost stable circular orbit (ISCO) for a Kerr black hole
was interrogated in \cite{8}. In \cite{9}, the researchers argued that the particle acceleration to swiftly huge energy is one of the global properties of an extremal Kerr black hole not only in astrophysics but also in more common circumstances. In \cite{10}, the author discussed the collision for two neutral particles in the vicinity of the near-horizon extremal Kerr black hole and exposed that the CME is finite for any permissible value of the particle parameters. The CME for two colliding general geodesic massive and massless particles around a Kerr black hole was accessed in \cite{11}. The authors discovered that, in the direct collision scenario, an arbitrarily huge CME can occur near the horizon of an extremal Kerr black hole at the equator and on a belt centered at the equator lies between latitudes $\pm a\cos(\sqrt{3}-1)\simeq \pm42.94^{\circ}$. In \cite{12}, the researcher elucidated on the feasibility of having infinite CME in the center of mass frame of colliding particles is a common characteristic of a Kerr black hole.

In \cite{13}, the authors inspected the collision of two general geodesic particles in the vicinity of a Kerr-Newman black hole and attain the CME of the non-marginally and marginally bound critical particles. In \cite{14}, the researchers investigated the CME over a Kerr-Newman black hole. They revealed that the unlimited CME needs three conditions: (a) the collision takes place at the horizon of an extremal black hole, (b) one of the colliding particles has critical angular momentum, and (c) the spin parameter $a$ permits $1/\sqrt{3}$ $\leq a\leq 1$. In \cite{15}, the author contemplated the collision for a freely falling neutral particle with a charged particle swirling in the circular orbit around a Schwarzschild black hole. In \cite{16}, the researchers discussed the collision for two particles with different rest masses moving in the equatorial plane of a Kerr-Taub-NUT black hole. They argued that the CME relies upon the spin parameter $a$ and NUT (Newman-Unti-Tamburino) charge $n$. In \cite{17,18,19}, the authors surveyed the Exact Lense-Thirring (LT) precession and spontaneous geodesics in the ISCO of a Kerr-Taub-NUT black hole. The CME of the collision for two uncharged particles falling freely from rest at infinity in the vicinity of a charged, rotating and accelerating black hole was discussed in \cite{20}. In \cite{21}, the authors investigated the CME of the collision for two neutral particles with different rest masses falling freely from rest at infinity around a Kerr-Newman-Taub-NUT (KNTN) black hole. In addition, they discussed the CME near the horizon(s) of an extremal and non-extremal KNTN black hole and demonstrated that the CME near the horizon(s) of an extremal and non-extremal KNTN black hole is arbitrarily high when the specific angular momentum of one of the colliding particles is equal to the critical angular momentum for a non-vanishing spin parameter $a$. They endorsed the Hamilton-Jacobi approach to study the dynamics of a neutral particle.

The collision for two particles in the vicinity of a charged black string was examined in \cite{22}. It was discovered that the CME is swiftly huge at the outer horizon if one of the colliding particles has critical charge. In \cite{23}, the researchers investigated the collision for two particles in the vicinity of a stringy black hole. They exhibited that the CME is swiftly huge under two conditions: (a) the spin parameter $a\neq0$, and (b) one of the colliding particles should have critical angular momentum. The CME in the absence and presence of a magnetic field in the vicinity of a Schwarzschild-like black hole was interrogated in \cite{24}. The particle acceleration mechanism in $S^{2}\times R^{1}$ topology, namely, in the spacetime of the five-dimensional compact black string, has been studied in \cite{25}. It was found that the scattering energy of particles in the center of mass frame can take arbitrarily large values for an extremal and non-extremal black string. In \cite{26}, the authors investigated the CME for two colliding neutral particles at the horizon of a slowly rotating black hole in the Horava-Lifshitz theory of gravity and a topological Lifshitz black hole and showed that the CME remains finite. In \cite{27}, the researchers examined the collision for test charged particles in the vicinity of the event horizon of a weakly magnetized static black hole with gravitomagnetic charge. The author of Ref.~\cite{28} argued that the BSW effect exists for a non-rotating but charged black hole even for the simplest case of radial motion of particles in a Reissner-Nordstr\"{o}m black hole. The CME of the collision for charged particles in a Bardeen black hole was inspected in \cite{29}. In \cite{30}, the researcher interrogated the effect of unbound acceleration of particles for Reissner-Nordstr\"{o}m and Kerr black holes. In \cite{31}, the researchers scrutinized the CME near the horizon of a non-extremal Plebanski-Demianski black hole without NUT parameter. The CME in the vicinity of Ay\`{o}n-Beato--Garc\`{i}a--Bronnikov (ABGB), Einstein-Maxwell-dilaton-axion (EMDA) and  Ba\~{n}ados-Teitelboim-Zanelli (BTZ) black holes was interrogated in \cite{32}.

In \cite{33}, the authors investigated the dynamics of a neutral and a charged particle around a black hole in modified gravity immersed in a uniform magnetic field. The authors considered the static, axially symmetric, rotating and charged Kerr-Sen Dilaton-Axion black hole metric in generalized Boyer-Lindquist coordinates as particle accelerators in \cite{34}. In \cite{35}, the authors studied the particle collisions within the context of $f(R)$ gravity described by $f(R)=R+2\alpha\sqrt{R}$, where $R$ stands for the Ricci scalar and $\alpha$ is a non-zero constant. In \cite{36}, the authors discussed the CME for two neutral particles with same rest masses falling from rest at infinity and colliding near the horizons of rotating modified Hayward and rotating modified Bardeen black holes. The authors discussed the collision of two particles where first particle comes from far to the outer horizon of the Reissner-Nordstr\"{o}m black hole and second particle emanates from the white hole region \cite{37}.

A vacuum solution of the modified gravitational field equations is a Kerr-Modified-Gravity (Kerr-MOG) black hole,
which in addition the spin parameter $a$ lifts the gravitational constant $G=(1+\alpha)G_{N}$. We follow the Hamilton-Jacobi approach to analysis the dynamics of a neutral particle in the vicinity of a Kerr-MOG black hole. We do not restrain the dynamics and collision to the equatorial plane only. Instead we designate arbitrary $\theta$ and fix $\theta=$ {\Large $\frac{\pi}{2}$} only as a exceptional case. We discuss the detailed behavior of the CME for three neutral particles with different rest masses $m_{1}$, $m_{2}$ and $m_{3}$ falling freely from rest at infinity in the background of a Kerr-MOG black hole. We determine the CME when the collision happens at some radial coordinate $r$ and angle $\theta$ nearby the horizon. We show that the CME near the horizon(s) of an extremal and non-extremal Kerr-MOG black hole is swiftly huge when the specific angular momentum of one of the colliding particles is equal to the critical angular momentum and non-vanishing spin parameter $a$.

The paper is established as follows. In Sec. II, we will discuss the equations of motion for a neutral particle in the background of a Kerr-Modified-Gravity (Kerr-MOG) black hole. In Sec. III, we will investigate the CME of the collision for three neutral particles and discuss the properties. In Sec. IV, we will illustrate a brief conclusion. We use the system of units $c=1$ all over this paper.
 \section{Equations of motion in the background of a Kerr-MOG black hole}
In this section, we will discuss the equations of motion for a neutral particle in the vicinity of a Kerr-MOG black hole. Let us first provide a concise analysis of a Kerr-MOG black hole. The Kerr-MOG black hole is a geometrically stationary and axisymmetric object, which is a valuable solution of vacuum field equations. The Kerr-MOG black hole is defined by the following parameters i.e., the mass $M$, spin parameter $a$ and gravitational constant $G=(1+\alpha)G_{N}$ where $\alpha$ determines the gravitational field strength and $G_{N}$ is Newton’s gravitational
constant. The Kerr-MOG black hole can be interpreted by the metric in the Boyer-Lindquist coordinates $(t, r, \theta, \phi)$ as in \cite{38}
\begin{eqnarray}\label{1}
ds^{2}&=&-\frac{1}{\Sigma}(\Delta-a^{2}\sin^{2}\theta)dt^{2}+\frac{2}{\Sigma}\big(\Delta-r^{2}-a^{2}\big)a\sin^{2}\theta dtd\phi+\frac{1}{\Sigma}\big((r^{2}+a^{2})^{2}-a^{2}\Delta\sin^{2}\theta\big)\sin^{2}\theta d\phi^{2}\notag\\&&
+\frac{\Sigma}{\Delta}dr^{2}+\Sigma d\theta^{2},
\end{eqnarray}
where $\Sigma$ and $\Delta$ are respectively designated by
\begin{eqnarray}\label{2}
\Sigma&=&r^{2}+a^{2}\cos^{2}\theta, \notag\\     \Delta &=&r^{2}+a^{2}-(2r-\alpha G_{N}M)GM.
\end{eqnarray}
The Kerr metric is acquired from the Kerr-MOG metric by setting $\alpha=0$. For $a=0$, we obtain metric for Schwarzschild-MOG black hole. The metric (\ref{1}) turns out to be singular if $\Sigma=0$ or $\Delta=0$, whereas $\Sigma=0$ is the curvature singularity and $\Delta=0$ is the coordinate singularity. Here, $\Sigma=0$ indicates $r=0$ and $\theta=$ {\Large $\frac{\pi}{2}$}. The horizon(s) of the Kerr-MOG black hole take place at $ r_{\pm}=GM\pm\sqrt{GG_{N}M^{2}-a^{2}}$,
where $r_{+}$ and $r_{-}$ give description of the outer and inner horizons, respectively, which are zeros of the polynomial $\Delta$. The presence of the horizons enforce $a^{2}\leq GG_{N}M^{2}$, where ``$=$" and ``$>$" agree with the extremal and non-extremal Kerr-MOG black holes, respectively.

Now, let us study the equations of motion for a neutral particle of mass $m$ in the vicinity of a Kerr-MOG black hole. The Lagrangian of the particle can be specified by
\begin{equation}\label{3}
 \mathcal{L}=\frac{1}{2}g_{\xi\eta}\dot{x}^{\xi}\dot{x}^{\eta},
\end{equation}
where the overdot designates differentiation with respect to an affine parameter $\lambda$. The relation between affine parameter $\lambda$ and proper time $\tau$ is $\tau=m\lambda$.
The normalization condition is {\Large $\frac{1}{m^2}$}$g_{\xi\eta}\dot{x}^{\xi}\dot{x}^{\eta}=\kappa$, where $\kappa=-1,0,1$ for timelike, null and spacelike geodesics, respectively. Timelike geodesic is followed by massive particle, so we consider $\kappa=-1$. The 4-momentum is given by
\begin{equation}\label{4}
  P_{\xi}=\frac{\partial \mathcal{L}}{\partial \dot{x}^{\xi}}=g_{\xi\eta}\dot{x}^{\eta}.
\end{equation}
The relation between 4-velocity and 4-momentum is given by
 \begin{equation}\label{5}
 u_{\xi}=\frac{P_{\xi}}{m},
\end{equation}
where $u^{\eta}=$ {\Large $\frac{dx^{\eta}}{d\tau}$}, $\tau$ is the proper time. Using Eq. (\ref{4}), we can show $\dot{x}^{\xi}$ in terms of inverse metric component and 4-momentum as $\dot{x}^{\xi}=g^{\xi\eta}P_{\eta}$. The Hamiltonian is defined by
\begin{equation}\label{6}
  \mathcal{H}=P_{\xi}\dot{x}^{\xi}-\mathcal{L}=\frac{1}{2}g^{\xi\eta}{P}_{\xi}{P}_{\eta},
\end{equation}
Moreover, the Hamilton-Jacobi equation is described by
\begin{equation}\label{7}
  \mathcal{H}=-\frac{\partial S}{\partial \lambda}=\frac{1}{2}g^{\xi\eta}\frac{\partial S}{\partial x^{\xi}}\frac{\partial S}{\partial x^{\eta}},
\end{equation}
where $S$ is named as Jacobi action and
\begin{equation}\label{8}
  \frac{\partial S}{\partial x^{\xi}}=P_{\xi}.
\end{equation}
The Hamilton-Jacobi equation admits separation of variables as
\begin{equation}\label{9}
S(\lambda,t,r,\theta,\phi) =\frac{1}{2}m^2\lambda-\mathcal{E}t+S_{r}(r)+S_{\theta}(\theta)+h\phi,
\end{equation}
where $\mathcal{E}$ and $h$ are energy and angular momentum of the particle, respectively. The functions
$S_{r}$ and $S_{\theta}$ are random functions of $r$ and $\theta$, respectively. Here, {\Large $\frac{1}{m}$}{\Large $\frac{\partial S}{\partial t}$} $=-E$ and
{\Large $\frac{1}{m}$}{\Large $\frac{\partial S}{\partial \phi}$} $=L$, where $E=${\Large $\frac{\mathcal{E}}{m}$}
and $L=${\Large $\frac{h}{m}$} are the specific energy and specific angular momentum of the particle. These relations and Eq. (\ref{8}) give
\begin{eqnarray}
 E&=&-\frac{P_{t}}{m}=\frac{1}{\Sigma}(\Delta-a^{2}\sin^{2}\theta)u^{t}-\frac{1}{\Sigma}\big(\Delta-r^{2}-a^{2}\big)a\sin^{2}\theta u^{\phi},\label{10}\\
 L&=&\frac{P_{\phi}}{m}=\frac{1}{\Sigma}\big(\Delta-r^{2}-a^{2}\big)a\sin^{2}\theta u^{t}+\frac{1}{\Sigma}\big((r^{2}+a^{2})^{2}-a^{2}\Delta\sin^{2}\theta\big)\sin^{2}\theta u^{\phi}.\label{11}
\end{eqnarray}
Work out with Eqs. (\ref{10}) and (\ref{11}), we get
\begin{eqnarray}
\Sigma u^{t}&=&a(L-aE\sin^{2}\theta)+\frac{r^{2}+a^{2}}{\Delta}[E(r^{2}+a^{2})-aL],\label{12}\\
\Sigma u^{\phi}&=&\frac{L}{\sin^{2}\theta}-aE+\frac{a}{\Delta}[E(r^{2}+a^{2})-aL].\label{13}
\end{eqnarray}
Eqs. (\ref{7}) and (\ref{9}) allow
\begin{eqnarray}\label{14}
-\frac{\Delta}{m^2}\bigg(\frac{\partial S_{r}}{\partial r}\bigg)^{2}-r^{2}-\big(L-aE\big)^{2}+\frac{1}{\Delta}\Big(\big(r^{2}+a^{2}\big) E-aL\Big)^{2}\notag\\=\frac{1}{m^2}\bigg(\frac{\partial S_{\theta}}{\partial \theta}\bigg)^{2}+\cos^{2}\theta\bigg(\big(1-E^{2}\big)a^{2}+\frac{L^{2}}{\sin^{2}\theta}\bigg).
\end{eqnarray}
The right-hand side of Eq. (\ref{14}) does not rely upon $r$ while the left-hand side does not rely upon $\theta$, hence each side must be stable. This is equal to conserved quantity and it is named as the Carter constant represented by $K$. Thus
\begin{equation}\label{15}
\frac{1}{m^2}\bigg(\frac{\partial S_{\theta}}{\partial \theta}\bigg)^{2}+\cos^{2}\theta\bigg(\big(1-E^{2}\big)a^{2}+\frac{L^{2}}{\sin^{2}\theta}\bigg)=K,
\end{equation}
\begin{equation}\label{16}
\frac{\Delta}{m^2}\bigg(\frac{\partial S_{r}}{\partial r}\bigg)^{2}+r^{2}+\big(L-aE\big)^{2}-\frac{1}{\Delta}\Big(\big(r^{2}+a^{2}\big) E-aL\Big)^{2}=-K.
\end{equation}
Using the results $u_{r}=$ {\Large $\frac{1}{m}$}{\Large $\frac{\partial S_{r}}{\partial r}$} and $u_{\theta}=$ {\Large $\frac{1}{m}$}{\Large $\frac{\partial S_{\theta}}{\partial \theta}$}, we have
\begin{eqnarray}
\Sigma u^{\theta}&=&\pm\sqrt{c},\label{17}\\
\Sigma u^{r}&=&\pm\sqrt{b},\label{18}
\end{eqnarray}
where
\begin{eqnarray}
c=c(\theta)&=&K-\cos^{2}\theta\bigg((1-E^{2})a^{2}+\frac{L^{2}}{\sin^{2}\theta}\bigg), \label{19}\\
b=b(r)&=&\big[E(r^{2}+a^{2})-aL\big]^{2}-\Delta\big[K+r^{2}+(L-aE)^{2}\big].\label{20}
\end{eqnarray}
The $\pm$ signs are self-sufficient and one must be persistent in that choice. The $+(-)$ sign agrees to the outgoing(ingoing) geodesics. The Carter constant $K$ disappears in the equatorial plane $\Big(\theta=$ {\Large $\frac{\pi}{2}$}$\Big)$. The radial equation of motion (\ref{18}) can also be composed as
\begin{equation}\label{21}
\frac{1}{2}(u^{r})^{2}+V_{\text{eff}}(r,\theta)=\frac{1}{2}(E^{2}-1),
\end{equation}
where the effective potential $V_{\text{eff}}(r,\theta)$ is
\begin{eqnarray}\label{22}
V_{\text{eff}}(r,\theta)&=&\frac{1}{2(r^{2}+a^{2}\cos^{2}\theta)^{2}}\Big[-\big(-a^{4}\cos^{4}\theta+a^{2}r^{2}
(\sin^{2}\theta-\cos^{2}\theta)+a^{2}\zeta\big)E^{2}+\big(r^{2}-\zeta\big)L^{2}\notag\\&&+2a\zeta LE+\big(r^{2}+a^{2}-\zeta\big)(K-a^{2}\cos^{2}\theta)+\big(a^{2}\sin^{2}\theta-\zeta\big)\big(r^{2}+a^{2}\cos^{2}\theta\big)\Big],
\end{eqnarray}
where $\zeta=(1+\alpha)(2r-\alpha G_{N}M)G_{N}M$.  Also for $r\rightarrow\infty$, $V_{\text{eff}}(r,\theta)\rightarrow0$. From Eqs. (\ref{18}) and (\ref{21}), we achieve that $V_{\text{eff}}(r,\theta)=$ {\Large $\frac{1}{2}$} $(E^2-1)-${\Large $\frac{b(r)}{2\Sigma^2}$}. Note that from Eqs. (\ref{17}) and (\ref{18}), $c\geq0$ and $b\geq0$ must be agreed for the admitted motion. Hence, the admitted and restricted regions for $V_{\text{eff}}(r,\theta)$ are represented by $V_{\text{eff}}(r,\theta)\leq${\Large $\frac{1}{2}$} $(E^2-1)$ and $V_{\text{eff}}(r,\theta)>${\Large $\frac{1}{2}$} $(E^2-1)$, respectively. The effective potential in the equatorial plane is obtained by
\begin{eqnarray}\label{23}
V_{\text{eff}}\Big(r,\frac{\pi}{2}\Big)&=&\frac{1}{2r^{4}}\Big[-a^{2}\big(r^{2}+\zeta\big)E^{2}+\big(r^{2}-\zeta\big)L^{2}+2a\zeta LE+r^{2}\big(a^{2}-\zeta\big)\Big].
\end{eqnarray}
The function $b(r)$ can also be composed as
\begin{eqnarray}\label{24}
b(r)&=&(E^{2}-1)r^{4}+2(1+\alpha)G_{N}Mr^{3}+[(E^{2}-1)a^{2}-L^{2}-K-\alpha(1+\alpha)G^{2}_{N}M^{2}]r^{2}\notag\\&&+2(1+\alpha)[(L-aE)^{2}+K]G_{N}Mr
-a^{2}K-\alpha(1+\alpha)[(L-aE)^{2}+K]G^{2}_{N}M^{2}.
\end{eqnarray}
Observe that coefficient of the highest power of $r$ on the right-hand side of Eq. (\ref{24}) is positive if $E>1$. Only in this case, the motion can be unbounded. For $E<1$, the motion is bounded i.e., the particle cannot approach the horizon(s) of the black hole. The motion is marginally bounded for $E=1$, i.e., the motion is either bounded or unbounded. In the case of $E=1$, the particle's motion relies upon the black hole parameters and specific angular momentum for the admitted and restricted regions of $b(r)$ and $c(\theta)$ but in the equatorial plane the motion can be fully examined by $b(r)$ or $V_{\text{eff}}\Big(r,$ {\Large $\frac{\pi}{2}$}$\Big)$. The description of a concepts bound, unbound and marginally bound particles are related to the particle whose motion is bounded, unbounded and marginally bounded. We get $V_{\text{eff}}(r,\theta)<0$ and $V_{\text{eff}}(r,\theta)\leq0$ for bound and marginally bound particles, respectively.

We need to enforce the ``forward-in-time" condition $u^{t}>0$ along the geodesics which demonstrates that the time coordinate $t$ increases along the trajectory of the particle's motion. This condition at $r\rightarrow r_{+}$ diminishes to
\begin{equation}\label{25}
L\leq\frac{\zeta_{+} E}{a},~~~~~ \text{where}~~~~~\zeta_{+}=(2r_{+}-\alpha G_{N}M)GM.
\end{equation}
Here, we obtain the upper bound of the specific angular momentum at the outer horizon of the non-extremal Kerr-MOG black hole which is called the critical angular momentum and is designated by $\hat{L}_{+}$ i.e.,
\begin{equation}\label{26}
\hat{L}_{+}=\frac{\zeta_{+} E}{a}.
\end{equation}
Likewise, the critical angular momentum at the inner horizon of the non-extremal Kerr-MOG black hole is represented by
\begin{equation}\label{27}
\hat{L}_{-}=\frac{\zeta_{-} E}{a},~~~~~ \text{where}~~~~~\zeta_{-}=(2r_{-}-\alpha G_{N}M)GM.
\end{equation}
For the extremal Kerr-MOG black hole, we have $a^{2}=(1+\alpha)G^{2}_{N}M^{2}$ in Eq. (\ref{26}), which shows the critical angular momentum at the horizon of the extremal Kerr-MOG black hole
\begin{equation}\label{28}
\hat{L}=\frac{(2+\alpha)GG_{N}M^{2}E}{a}.
\end{equation}
For $a=0$, Eqs. (\ref{26}), (\ref{27}) and (\ref{28}) become undefined, so we will consider $a\neq0$ all over in this paper.

\section{Center of mass energy for three particles}
 In this section, we will analyse the CME of the collision for three neutral particles $(1, 2$ and $3)$ with different rest masses $m_{1}$, $m_{2}$ and $m_{3}$ falling freely from rest at infinity towards a Kerr-MOG black hole. Let us assume that these particles collide at some radial coordinate $r$ which are not restrained in the equatorial plane. The CME of the collision is obtained by
\begin{equation}\label{34}
E_{\text{cm}}=\sqrt{m_{1}^{2}+m_{2}^{2}+m_{3}^{2}-2m_{1}m_{2}g_{\mu\nu}u_{(1)}^{\mu}u_{(2)}^{\nu}
-2m_{1}m_{3}g_{\mu\nu}u_{(1)}^{\mu}u_{(3)}^{\nu}-2m_{2}m_{3}g_{\mu\nu}u_{(2)}^{\mu}u_{(3)}^{\nu}}.
\end{equation}
For the Kerr-MOG metric (\ref{1}), using Eqs. (\ref{12}), (\ref{13}), (\ref{17}) and (\ref{18}) into Eq. (\ref{34}), we have
\begin{eqnarray}\label{35}
E_{\text{cm}}&=&\Bigg(m_{1}^{2}+m_{2}^{2}+m_{3}^{2}+2m_{1}m_{2}\frac{A(r,\theta)-B(r,\theta)-C(r,\theta)}{D(r,\theta)}
+2m_{1}m_{3}\frac{H(r,\theta)-I(r,\theta)-J(r,\theta)}{D(r,\theta)}\notag\\&&
+2m_{2}m_{3}\frac{X(r,\theta)-Y(r,\theta)-Z(r,\theta)}{D(r,\theta)}\Bigg)^{\frac{1}{2}},
\end{eqnarray}
where $A(r, \theta)$, $B(r,\theta)$, $C(r,\theta)$, $D(r,\theta)$, $H(r,\theta)$, $I(r,\theta)$, $J(r,\theta)$, $X(r,\theta)$, $Y(r,\theta)$ and $Z(r,\theta)$ are obtained by
\begin{eqnarray}\label{36}
A(r,\theta)&=&(a^{2}\sin^{2}\theta-\Delta)L_{1}L_{2}+\big((r^{2}+a^{2})^{2}-a^{2}\Delta\sin^{2}\theta\big)\sin^{2}\theta E_{1}E_{2}+\sin^{2}\theta\big(a\Delta\notag\\&&-a(r^{2}+a^{2})\big)\big(L_{1}E_{2}+L_{2}E_{1}\big),\notag\\
B(r,\theta)&=&\sin^{2}\theta\sqrt{b_{1}(r)b_{2}(r)},\notag\\
C(r,\theta)&=&\Delta\sin^{2}\theta\sqrt{c_{1}(\theta)c_{2}(\theta)},\notag\\
D(r,\theta)&=&\Delta\Sigma\sin^{2}\theta,\notag\\
H(r,\theta)&=&(a^{2}\sin^{2}\theta-\Delta)L_{1}L_{3}+\big((r^{2}+a^{2})^{2}-a^{2}\Delta\sin^{2}\theta\big)\sin^{2}\theta E_{1}E_{3}+\sin^{2}\theta\big(a\Delta\notag\\&&\allowdisplaybreaks-a(r^{2}+a^{2})\big)\big(L_{1}E_{3}+L_{3}E_{1}\big),\notag\\
I(r,\theta)&=&\sin^{2}\theta\sqrt{b_{1}(r)b_{3}(r)},\\
J(r,\theta)&=&\Delta\sin^{2}\theta\sqrt{c_{1}(\theta)c_{3}(\theta)},\notag\\
X(r,\theta)&=&(a^{2}\sin^{2}\theta-\Delta)L_{2}L_{3}+\big((r^{2}+a^{2})^{2}-a^{2}\Delta\sin^{2}\theta\big)\sin^{2}\theta E_{2}E_{3}+\sin^{2}\theta\big(a\Delta\notag\\&&-a(r^{2}+a^{2})\big)\big(L_{2}E_{3}+L_{3}E_{2}\big),\notag\\
Y(r,\theta)&=&\sin^{2}\theta\sqrt{b_{2}(r)b_{3}(r)},\notag\\
Z(r,\theta)&=&\Delta\sin^{2}\theta\sqrt{c_{2}(\theta)c_{3}(\theta)},\notag\\
b_{i}(r)&=&\big[\big(r^{2}+a^{2}\big)E_{i}-aL_{i}\big]^{2}-\Delta\big[K_{i}+r^{2}+(L_{i}-aE_{i})^{2}\big],\notag\\
c_{i}(\theta)&=&K_{i}-\cos^{2}\theta\bigg(\big(1-E_{i}^{2}\big)a^{2}+\frac{L_{i}^{2}}{\sin^{2}\theta}\bigg).\notag
\end{eqnarray}
At this place, $K_{i}$, $E_{i}$ and $L_{i}$  are respectively the Carter constant, specific energy and specific angular momentum of the $i$th particle. Clearly, the CME (\ref{35}) is undeviating under the exchange of the quantities $m_{1}\leftrightarrow m_{2}$, $E_{1}\leftrightarrow E_{2}$ and $L_{1}\leftrightarrow L_{2}$. For $\theta=\frac{\pi}{2}$ and $m_{3}=0$, Eq. (\ref{35}) diminishes to the outcome obtained in \cite{39}.
\subsection{Near-horizon collision of particles around the non-extremal Kerr-MOG black hole}
Let us investigate the properties of the CME (\ref{35}) as the particles access the horizons $r_{+}$ and $r_{-}$ of the non-extremal Kerr-MOG black hole.
\subsubsection{Collision at the outer horizon}
The terms $\frac{A(r, \theta)-B(r,\theta)-C(r,\theta)}{D(r,\theta)}$, $\frac{H(r, \theta)-I(r,\theta)-J(r,\theta)}{D(r,\theta)}$  and $\frac{X(r, \theta)-Y(r,\theta)-Z(r,\theta)}{D(r,\theta)}$ of right-hand side of Eq. (\ref{35}) become $\big(\frac{0}{0}\big)$ at $r_{+}$. Utilizing L'Hospital's rule and the identity $r_{+}^{2}+a^{2}-(1+\alpha)(2r_{+}-\alpha G_{N}M)G_{N}M=0$, the value of the CME at $r_{+}$ evolves into
\begin{eqnarray}\label{37}
E_{\text{cm}}\Big|_{r\rightarrow r_{+}}&=&\Bigg(m_{1}^{2}+m_{2}^{2}+m_{3}^{2}+\frac{2m_{1}m_{2}}{\partial_{r}D(r,\theta)}\big(\partial_{r}A(r,\theta)
-\partial_{r}B(r,\theta)-\partial_{r}C(r,\theta)\big)+
\frac{2m_{1}m_{3}}{\partial_{r}D(r,\theta)}\notag\\&&\times\big(\partial_{r}H(r,\theta)-\partial_{r}I(r,\theta)
-\partial_{r}J(r,\theta)\big)
+\frac{2m_{2}m_{3}}{\partial_{r}D(r,\theta)}\big(\partial_{r}X(r,\theta)-\partial_{r}Y(r,\theta)
\notag\\&&-\partial_{r}Z(r,\theta)\big)\Bigg)^{\frac{1}{2}}\Bigg|_{r\rightarrow r_{+}},
\end{eqnarray}
where
\begin{eqnarray}\label{38}
\partial_{r}A(r,\theta)\big|_{r\rightarrow r_{+}}&=&-2(r_{+}-GM)L_{1}L_{2}
+\big[4r_{+}\big(r_{+}^{2}+a^{2}\big)-2(r_{+}-GM)a^{2}\sin^{2}\theta\big]\sin^{2}\theta E_{1}E_{2}\notag\\&&
-2aGM\sin^{2}\theta\big(L_{1}E_{2}+L_{2}E_{1}\big),\notag\\
\partial_{r}B(r,\theta)\big|_{r\rightarrow r_{+}}&=&\bigg[\frac{\sin^{2}\theta}{2\sqrt{b_{1}(r)b_{2}(r)}}\bigg(b_{2}(r) \partial_{r}b_{1}(r)+b_{1}(r)\partial_{r} b_{2}(r)\bigg)\bigg]\bigg|_{r\rightarrow r_{+}},\notag\\
\partial_{r}C(r,\theta)\big|_{r\rightarrow r_{+}}&=&2(r_{+}-GM)\sin^{2}\theta\sqrt{c_{1}(\theta)c_{2}(\theta)},\notag\\
\partial_{r}D(r,\theta)\big|_{r\rightarrow r_{+}}&=&2(r_{+}-GM)\big(r_{+}^{2}+a^{2}\cos\theta^{2}\big)\sin^{2}\theta,\notag\\
\partial_{r}H(r,\theta)\big|_{r\rightarrow r_{+}}&=&-2(r_{+}-GM)L_{1}L_{3}
+\big[4r_{+}\big(r_{+}^{2}+a^{2}\big)-2(r_{+}-GM)a^{2}\sin^{2}\theta\big]\sin^{2}\theta E_{1}E_{3}\notag\\&&
-2aGM\sin^{2}\theta\big(L_{1}E_{3}+L_{3}E_{1}\big),\\
\partial_{r}I(r,\theta)\big|_{r\rightarrow r_{+}}&=&\bigg[\frac{\sin^{2}\theta}{2\sqrt{b_{1}(r)b_{3}(r)}}\bigg(b_{3}(r) \partial_{r}b_{1}(r)+b_{1}(r)\partial_{r} b_{3}(r)\bigg)\bigg]\bigg|_{r\rightarrow r_{+}},\notag\allowdisplaybreaks\\
\partial_{r}J(r,\theta)\big|_{r\rightarrow r_{+}}&=&2(r_{+}-GM)\sin^{2}\theta\sqrt{c_{1}(\theta)c_{3}(\theta)},\notag\allowdisplaybreaks\\
\partial_{r}X(r,\theta)\big|_{r\rightarrow r_{+}}&=&-2(r_{+}-GM)L_{2}L_{3}
+\big[4r_{+}\big(r_{+}^{2}+a^{2}\big)-2(r_{+}-GM)a^{2}\sin^{2}\theta\big]\sin^{2}\theta E_{2}E_{3}\notag\\&&
-2aGM\sin^{2}\theta\big(L_{2}E_{3}+L_{3}E_{2}\big),\notag\\
\partial_{r}Y(r,\theta)\big|_{r\rightarrow r_{+}}&=&\bigg[\frac{\sin^{2}\theta}{2\sqrt{b_{2}(r)b_{3}(r)}}\bigg(b_{3}(r) \partial_{r}b_{2}(r)+b_{2}(r)\partial_{r} b_{3}(r)\bigg)\bigg]\bigg|_{r\rightarrow r_{+}},\notag\\
\partial_{r}Z(r,\theta)\big|_{r\rightarrow r_{+}}&=&2(r_{+}-GM)\sin^{2}\theta\sqrt{c_{2}(\theta)c_{3}(\theta)},\notag\\
\partial_{r}b_{i}(r)\big|_{r\rightarrow r_{+}}&=&4r_{+}\big[\big(r_{+}^{2}+a^{2}\big)E_{i}-aL_{i}\big]E_{i}-2\big(r_{+}
-GM\big)\big[K_{i}+r_{+}^{2}+(L_{i}-aE_{i})^{2}\big].\notag
\end{eqnarray}
After elucidation, we achieve the CME at the outer horizon
\begin{eqnarray}\label{39}
E_{\text{cm}}\Big|_{r\rightarrow r_{+}}&=&\Bigg[m_{1}^{2}+m_{2}^{2}+m_{3}^{2}+2m_{1}m_{2}\Bigg(1+\frac{1}{2(\hat{L}_{+1}-
L_{1})(\hat{L}_{+2}-L_{2})}\bigg[\big[(\hat{L}_{+1}-L_{1})-(\hat{L}_{+2}-L_{2})\big]^{2}\notag\\&&
+\frac{1}{r_{+}^{2}+a^{2}\cos^{2}\theta}\bigg(\frac{r_{+}^{4}}{(r_{+}^{2}+a^{2})^{2}}
\big(L_{1}\hat{L}_{+2}-L_{2}\hat{L}_{+1}\big)^{2}+K_{2}(\hat{L}_{+1}-L_{1})^{2}+K_{1}(\hat{L}_{+2}
-L_{2})^{2}\notag\\&&-a^{2}\cos^{2}\theta\big[(\hat{L}_{+1}-L_{1})^{2}+(\hat{L}_{+2}-L_{2})^{2}\big]\bigg)\bigg]-
\frac{1}{r_{+}^{2}+a^{2}\cos^{2}\theta}\bigg[\cot^{2}\theta L_{1}L_{2}+\frac{a^{4}\cos^{2}\theta}
{(r_{+}^{2}+a^{2})^{2}}\notag\\&&\times\hat{L}_{+1}\hat{L}_{+2}+\sqrt{c_{1}(\theta)c_{2}(\theta)}\bigg]\Bigg)+2m_{1}m_{3}
\Bigg(1+\frac{1}{2(\hat{L}_{+1}-
L_{1})(\hat{L}_{+3}-L_{3})}\bigg[\big[(\hat{L}_{+1}-L_{1})\notag\\&&-(\hat{L}_{+3}-L_{3})\big]^{2}
+\frac{1}{r_{+}^{2}+a^{2}\cos^{2}\theta}\bigg(\frac{r_{+}^{4}}{(r_{+}^{2}+a^{2})^{2}}
\big(L_{1}\hat{L}_{+3}-L_{3}\hat{L}_{+1}\big)^{2}+K_{3}(\hat{L}_{+1}-L_{1})^{2}\notag\\&&+K_{1}(\hat{L}_{+3}
-L_{3})^{2}-a^{2}\cos^{2}\theta\big[(\hat{L}_{+1}-L_{1})^{2}+(\hat{L}_{+3}-L_{3})^{2}\big]\bigg)\bigg]-
\frac{1}{r_{+}^{2}+a^{2}\cos^{2}\theta}\bigg[\cot^{2}\theta\notag\\&&\times L_{1}L_{3}+\frac{a^{4}\cos^{2}\theta}
{(r_{+}^{2}+a^{2})^{2}}\hat{L}_{+1}\hat{L}_{+3}+\sqrt{c_{1}(\theta)c_{3}(\theta)}\bigg]\Bigg)
+2m_{2}m_{3}\Bigg(1+\frac{1}{2(\hat{L}_{+2}-
L_{2})(\hat{L}_{+3}-L_{3})}\notag\\&&\times\bigg[\big[(\hat{L}_{+2}-L_{2})-(\hat{L}_{+3}-L_{3})\big]^{2}
+\frac{1}{r_{+}^{2}+a^{2}\cos^{2}\theta}\bigg(\frac{r_{+}^{4}}{(r_{+}^{2}+a^{2})^{2}}
\big(L_{2}\hat{L}_{+3}-L_{3}\hat{L}_{+2}\big)^{2}\notag\\&&+K_{3}(\hat{L}_{+2}-L_{2})^{2}+K_{2}(\hat{L}_{+3}
-L_{3})^{2}-a^{2}\cos^{2}\theta\big[(\hat{L}_{+2}-L_{2})^{2}+(\hat{L}_{+3}-L_{3})^{2}\big]\bigg)\bigg]\notag\\&&-
\frac{1}{r_{+}^{2}+a^{2}\cos^{2}\theta}\bigg[\cot^{2}\theta L_{2}L_{3}+\frac{a^{4}\cos^{2}\theta}
{(r_{+}^{2}+a^{2})^{2}}\hat{L}_{+2}\hat{L}_{+3}+\sqrt{c_{2}(\theta)c_{3}(\theta)}\bigg]\Bigg)\Bigg]^{\frac{1}{2}},
\end{eqnarray}
where $\hat{L}_{+i}$ is the critical angular momentum for the $i$th particle, and can be drafted as $\hat{L}_{+i}=\frac{\zeta_{+} E_{i}}{a}$, where $\zeta_{+}=(2r_{+}-\alpha G_{N}M)GM$. $L_{i}=\hat{L}_{+i}$ is the essential condition to attain an arbitrarily huge CME. Selecting $E_{1}=E_{2}=E_{3}=E$, we achieve $\hat{L}_{+1}=\hat{L}_{+2}=\hat{L}_{+3}=\hat{L}_{+}$, and Eq. (\ref{39}) permits
\begin{eqnarray}\label{40}
E_{\text{cm}}\Big|_{r\rightarrow r_{+}}&=&\Bigg[m_{1}^{2}+m_{2}^{2}+m_{3}^{2}+2m_{1}m_{2}\Bigg(1+\frac{1}{2(\hat{L}_{+}-L_{1})(\hat{L}_{+}
-L_{2})}\bigg[(L_{1}-L_{2})^{2}
+\frac{1}{r_{+}^{2}+a^{2}\cos^{2}\theta}\notag\\&&\allowdisplaybreaks\times\bigg(\frac{r_{+}^{4}}{a^{2}}
E^{2}\big(L_{1}-L_{2}\big)^{2}+K_{2}(\hat{L}_{+}-L_{1})^{2}+K_{1}(\hat{L}_{+}-L_{2})^{2}-a^{2}\cos^{2}\theta
\big[(\hat{L}_{+}-L_{1})^{2}\notag\allowdisplaybreaks\\&&+(\hat{L}_{+}-L_{2})^{2}\big]\bigg)\bigg]-
\frac{1}{r_{+}^{2}+a^{2}\cos^{2}\theta}\bigg[\cot^{2}\theta L_{1}L_{2}
+a^{2}E^{2}\cos^{2}\theta
+\sqrt{c_{1}(\theta)c_{2}(\theta)}\bigg]\Bigg)\notag\allowdisplaybreaks\\&&+2m_{1}m_{3}\Bigg(1+\frac{1}{2(\hat{L}_{+}-L_{1})(\hat{L}_{+
}-L_{3})}\bigg[(L_{1}-L_{3})^{2}
+\frac{1}{r_{+}^{2}+a^{2}\cos^{2}\theta}\bigg(\frac{r_{+}^{4}}{a^{2}}
E^{2}\big(L_{1}-L_{3}\big)^{2}\notag\\&&+K_{3}(\hat{L}_{+}-L_{1})^{2}+K_{1}(\hat{L}_{+}-L_{3})^{2}-a^{2}\cos^{2}\theta
\big[(\hat{L}_{+}-L_{1})^{2}+(\hat{L}_{+}-L_{3})^{2}\big]\bigg)\bigg]\notag\\&&\allowdisplaybreaks-
\frac{1}{r_{+}^{2}+a^{2}\cos^{2}\theta}\bigg[\cot^{2}\theta L_{1}L_{3}
+a^{2}E^{2}\cos^{2}\theta
+\sqrt{c_{1}(\theta)c_{3}(\theta)}\bigg]\Bigg)+2m_{2}m_{3}\Bigg(1\notag\\&&+\frac{1}{2(\hat{L}_{+}-L_{2})(\hat{L}_{+}
-L_{3})}\bigg[(L_{2}-L_{3})^{2}
+\frac{1}{r_{+}^{2}+a^{2}\cos^{2}\theta}\bigg(\frac{r_{+}^{4}}{a^{2}}
E^{2}\big(L_{2}-L_{3}\big)^{2}+K_{3}(\hat{L}_{+}\notag\\&&-L_{2})^{2}+K_{2}(\hat{L}_{+}-L_{3})^{2}-a^{2}\cos^{2}\theta
\big[(\hat{L}_{+}-L_{2})^{2}+(\hat{L}_{+}-L_{3})^{2}\big]\bigg)\bigg]-
\frac{1}{r_{+}^{2}+a^{2}\cos^{2}\theta}\notag\\&&\times\bigg[\cot^{2}\theta L_{2}L_{3}
+a^{2}E^{2}\cos^{2}\theta
+\sqrt{c_{2}(\theta)c_{3}(\theta)}\bigg]\Bigg)\Bigg]^{\frac{1}{2}}.
\end{eqnarray}

Let us examine a marginally bound particle $(E=1)$ with the critical angular momentum $\hat{L}_{+}$. The conditions for the admitted region, $b(r)\geq0$ and $c(\theta)\geq0$ commit the upper and lower bounds for the Carter constant $K$ stated below
\begin{equation}\label{41}
\cot^{2}\theta\frac{(r_{+}^{2}+a^{2})^{2}}{a^{2}}\leq K\leq \frac{(r+r_{+})^{2}(r-r_{+})}{r-r_{-}}-r^{2}-\frac{r_{+}^{4}}{a^{2}}.
\end{equation}
Deal with Eq. (\ref{41}), we discover the condition for the marginally bound particle with the critical angular momentum to approach the outer horizon of the non-extremal Kerr-MOG black hole
\begin{equation}\label{44}
\Big(\frac{(r+r_{+})^{2}(r-r_{+})}{r-r_{-}}-r^{2}+a^{2}+2r_{+}^{2}\Big)\cos^{2}\theta-\frac{(r+r_{+})^{2}(r-r_{+})}{r-r_{-}}+r^{2}
+\frac{r_{+}^{4}}{a^{2}}\leq0,~~~~\text{for~any}~~~~r\geq r_{+}.
\end{equation}

If one prefers $\theta=$ {\Large $\frac{\pi}{2}$}, the CME (\ref{39}) at the outer horizon of the non-extremal Kerr-MOG black hole refers to
\begin{eqnarray}\label{45}
E_{\text{cm}}\Big|_{r\rightarrow r_{+}}&=&\Bigg[m_{1}^{2}+m_{2}^{2}+m_{3}^{2}+2m_{1}m_{2}\Bigg(1+\frac{1}{2(\hat{L}_{+1}-
L_{1})(\hat{L}_{+2}-L_{2})}\bigg(\big[(\hat{L}_{+1}-L_{1})-(\hat{L}_{+2}-L_{2})\big]^{2}\notag\\&&\allowdisplaybreaks
+\frac{r_{+}^{2}}{(r_{+}^{2}+a^{2})^{2}}
\big(L_{1}\hat{L}_{+2}-L_{2}\hat{L}_{+1}\big)^{2}\bigg)\Bigg)+2m_{1}m_{3}
\Bigg(1+\frac{1}{2(\hat{L}_{+1}-
L_{1})(\hat{L}_{+3}-L_{3})}\bigg(\big[(\hat{L}_{+1}\notag\\&&-L_{1})-(\hat{L}_{+3}-L_{3})\big]^{2}
+\frac{r_{+}^{2}}{(r_{+}^{2}+a^{2})^{2}}
\big(L_{1}\hat{L}_{+3}-L_{3}\hat{L}_{+1}\big)^{2}\bigg)\Bigg)
+2m_{2}m_{3}\Bigg(1\notag\\&&+\frac{1}{2(\hat{L}_{+2}-
L_{2})(\hat{L}_{+3}-L_{3})}\bigg(\big[(\hat{L}_{+2}-L_{2})-(\hat{L}_{+3}-L_{3})\big]^{2}
+\frac{r_{+}^{2}}{(r_{+}^{2}+a^{2})^{2}}
\big(L_{2}\hat{L}_{+3}\notag\\&&-L_{3}\hat{L}_{+2}\big)^{2}\bigg)\Bigg)\Bigg]^{\frac{1}{2}},
\end{eqnarray}
which is undoubtedly finite for all values of $L_{1}$, $L_{2}$ and $L_{3}$ except when $L_{1}$, $L_{2}$ or $L_{3}$ is approximately equal to the critical angular momentum $\hat{L}_{+i}$, for which the neutral particles collide with an arbitrarily huge CME. In the case of the identical specific energies, the arrangement of the CME (\ref{45}) at $r_{+}$ diminishes to
\begin{eqnarray}\label{46}
E_{\text{cm}}\Big|_{r\rightarrow r_{+}}&=&\Bigg[m_{1}^{2}+m_{2}^{2}+m_{3}^{2}+2m_{1}m_{2}\Bigg(1+\frac{1}{2(\hat{L}_{+}-L_{1})(\hat{L}_{+}
-L_{2})}\bigg((L_{1}-L_{2})^{2}
+\frac{r_{+}^{2}}{a^{2}}
E^{2}\big(L_{1}\notag\\&&\allowdisplaybreaks-L_{2}\big)^{2}\bigg)\Bigg)+2m_{1}m_{3}\Bigg(1+\frac{1}{2(\hat{L}_{+}-L_{1})(\hat{L}_{+
}-L_{3})}\bigg((L_{1}-L_{3})^{2}
+\frac{r_{+}^{2}}{a^{2}}
E^{2}\big(L_{1}-L_{3}\big)^{2}\bigg)\Bigg)\notag\\&&+2m_{2}m_{3}\Bigg(1
+\frac{1}{2(\hat{L}_{+}-L_{2})(\hat{L}_{+}
-L_{3})}\bigg((L_{2}-L_{3})^{2}
+\frac{r_{+}^{2}}{a^{2}}
E^{2}\big(L_{2}-L_{3}\big)^{2}\bigg)\Bigg)\Bigg]^{\frac{1}{2}}.
\end{eqnarray}
The largest and smallest value of specific angular momentum can be acquired by the equations
\begin{equation}\label{66}
V_{\text{eff}}\Big(r,\frac{\pi}{2}\Big)=0,~~~~\partial_{r}V_{\text{eff}}\Big(r,\frac{\pi}{2}\Big)=0.
\end{equation}
Then the interval $L\in\big[L_{\text{min}}, L_{\text{max}}\big]$ can be concluded from the above two equations. The intervals for the specific angular momentum for different values of $a$ and $\alpha$ are represented in Table ~\ref{KNTNtab:1}. Note that, with the increase of $\alpha$, the interval $L\in[L_{\text{min}}, L_{\text{max}}]$ becomes vast but with the increase of $a$, the interval $L\in[L_{\text{min}}, L_{\text{max}}]$ becomes slightly slim.
\begin{table}[h!]
\caption {The interval $L\in[L_{\text{min}}, L_{\text{max}}]$ with different spin parameter $a$ and gravitational field strength $\alpha$ for the non-extremal Kerr-MOG black hole.} \label{KNTNtab:1}
\begin{tabular}{|C{0.8cm}|C{3.75cm}|C{3.75cm}|C{3.75cm}|C{3.75cm}|}
 \hline
$\alpha$ & $a=0.2$ & $a=0.4$ & $a=0.6$ & $a=0.8$ \\
\hline\hline
$0.2$  & $\big[-~5.85750, ~~5.44138\big]$ & $\big[-~6.04509, ~~5.20554\big]$ & $\big[-~6.22216, ~~4.94287\big]$ & $\big[-~6.39031, ~~4.64139\big]$   \\ \hline
$0.4$ & $\big[-~7.83738, ~~7.41390\big]$ & $\big[-~8.03277, ~~7.18156\big]$ & $\big[-~8.21927, ~~6.93130\big]$ & $\big[-~8.39801, ~~6.65818\big]$   \\ \hline
$0.6$ & $\big[-10.13430, ~~9.70722\big]$ & $\big[-10.33470, ~~9.47798\big]$ & $\big[-10.52760, ~~8.31867\big]$ & $\big[-10.71390, ~~8.97869\big]$   \\ \hline
$0.8$ & $\big[-12.75000, ~12.3215\big]$ & $\big[-12.95360, ~12.09490\big]$ & $\big[-13.15100, ~11.85880\big]$ & $\big[-13.34270, ~11.61170\big]$   \\ \hline
$1$ & $\big[-15.68540, ~15.2565\big]$ & $\big[-15.89110, ~15.03230\big]$ & $\big[-16.09150, ~14.80060\big]$ & $\big[-16.28710, ~14.56070\big]$   \\ \hline
\end{tabular}
\end{table}
\subsubsection{Collision at the inner horizon}
Similarly the terms $\frac{A(r, \theta)-B(r,\theta)-C(r,\theta)}{D(r,\theta)}$, $\frac{H(r, \theta)-I(r,\theta)-J(r,\theta)}{D(r,\theta)}$  and $\frac{X(r, \theta)-Y(r,\theta)-Z(r,\theta)}{D(r,\theta)}$ of right-hand side of Eq. (\ref{35}) also turn into $\big(\frac{0}{0}\big)$ at $r_{-}$. By applying L'Hospital's rule we attain the CME for the three neutral particles at the inner horizon
\begin{eqnarray}\label{47}
E_{\text{cm}}\Big|_{r\rightarrow r_{-}}&=&\Bigg[m_{1}^{2}+m_{2}^{2}+m_{3}^{2}+2m_{1}m_{2}\Bigg(1+\frac{1}{2(\hat{L}_{-1}-
L_{1})(\hat{L}_{-2}-L_{2})}\bigg[\big[(\hat{L}_{-1}-L_{1})-(\hat{L}_{-2}-L_{2})\big]^{2}\notag\allowdisplaybreaks\\&&
+\frac{1}{r_{-}^{2}+a^{2}\cos^{2}\theta}\bigg(\frac{r_{-}^{4}}{(r_{-}^{2}+a^{2})^{2}}
\big(L_{1}\hat{L}_{-2}-L_{2}\hat{L}_{-1}\big)^{2}+K_{2}(\hat{L}_{-1}-L_{1})^{2}+K_{1}(\hat{L}_{-2}
-L_{2})^{2}\notag\allowdisplaybreaks\\&&-a^{2}\cos^{2}\theta\big[(\hat{L}_{-1}-L_{1})^{2}+(\hat{L}_{-2}-L_{2})^{2}\big]\bigg)\bigg]-
\frac{1}{r_{-}^{2}+a^{2}\cos^{2}\theta}\bigg[\cot^{2}\theta L_{1}L_{2}+\frac{a^{4}\cos^{2}\theta}
{(r_{-}^{2}+a^{2})^{2}}\notag\allowdisplaybreaks\\&&\times\hat{L}_{-1}\hat{L}_{-2}+\sqrt{c_{1}(\theta)c_{2}(\theta)}\bigg]\Bigg)+2m_{1}m_{3}
\Bigg(1+\frac{1}{2(\hat{L}_{-1}-
L_{1})(\hat{L}_{-3}-L_{3})}\bigg[\big[(\hat{L}_{-1}-L_{1})\notag\\&&\allowdisplaybreaks-(\hat{L}_{-3}-L_{3})\big]^{2}
+\frac{1}{r_{-}^{2}+a^{2}\cos^{2}\theta}\bigg(\frac{r_{-}^{4}}{(r_{-}^{2}+a^{2})^{2}}
\big(L_{1}\hat{L}_{-3}-L_{3}\hat{L}_{-1}\big)^{2}+K_{3}(\hat{L}_{-1}-L_{1})^{2}\notag\\&&+K_{1}(\hat{L}_{-3}
-L_{3})^{2}-a^{2}\cos^{2}\theta\big[(\hat{L}_{-1}-L_{1})^{2}+(\hat{L}_{-3}-L_{3})^{2}\big]\bigg)\bigg]-
\frac{1}{r_{-}^{2}+a^{2}\cos^{2}\theta}\bigg[\cot^{2}\theta\notag\\&&\times L_{1}L_{3}+\frac{a^{4}\cos^{2}\theta}
{(r_{-}^{2}+a^{2})^{2}}\hat{L}_{-1}\hat{L}_{-3}+\sqrt{c_{1}(\theta)c_{3}(\theta)}\bigg]\Bigg)
+2m_{2}m_{3}\Bigg(1+\frac{1}{2(\hat{L}_{-2}-
L_{2})(\hat{L}_{-3}-L_{3})}\notag\\&&\allowdisplaybreaks\times\bigg[\big[(\hat{L}_{-2}-L_{2})-(\hat{L}_{-3}-L_{3})\big]^{2}
+\frac{1}{r_{-}^{2}+a^{2}\cos^{2}\theta}\bigg(\frac{r_{-}^{4}}{(r_{-}^{2}+a^{2})^{2}}
\big(L_{2}\hat{L}_{-3}-L_{3}\hat{L}_{-2}\big)^{2}\notag\\&&+K_{3}(\hat{L}_{-2}-L_{2})^{2}+K_{2}(\hat{L}_{-3}
-L_{3})^{2}-a^{2}\cos^{2}\theta\big[(\hat{L}_{-2}-L_{2})^{2}+(\hat{L}_{-3}-L_{3})^{2}\big]\bigg)\bigg]\notag\\&&-
\frac{1}{r_{-}^{2}+a^{2}\cos^{2}\theta}\bigg[\cot^{2}\theta L_{2}L_{3}+\frac{a^{4}\cos^{2}\theta}
{(r_{-}^{2}+a^{2})^{2}}\hat{L}_{-2}\hat{L}_{-3}+\sqrt{c_{2}(\theta)c_{3}(\theta)}\bigg]\Bigg)\Bigg]^{\frac{1}{2}},
\end{eqnarray}
where $\hat L_{-i}$ is the critical angular momentum at the inner horizon, which can be drafted as $\hat{L}_{-i}=\frac{\zeta_{-} E_{i}}{a}$, where $\zeta_{-}=(2r_{-}-\alpha G_{N}M)GM$. An arbitrary huge CME can be achieved by selecting the condition $L_{i}=\hat{L}_{-i}$ for any of the three particles. The critical angular momentum is identical when both particles have the identical specific energy and is given by $\hat{L}_{-1}=\hat{L}_{-2}=\hat{L}_{-3}=\hat{L}_{-}$, thus the CME (\ref{47}) turns into
\begin{eqnarray}\label{48}
E_{\text{cm}}\Big|_{r\rightarrow r_{-}}&=&\Bigg[m_{1}^{2}+m_{2}^{2}+m_{3}^{2}+2m_{1}m_{2}\Bigg(1+\frac{1}{2(\hat{L}_{-}-L_{1})(\hat{L}_{-}
-L_{2})}\bigg[(L_{1}-L_{2})^{2}
+\frac{1}{r_{-}^{2}+a^{2}\cos^{2}\theta}\notag\\&&\times\bigg(\frac{r_{-}^{4}}{a^{2}}
E^{2}\big(L_{1}-L_{2}\big)^{2}+K_{2}(\hat{L}_{-}-L_{1})^{2}+K_{1}(\hat{L}_{-}-L_{2})^{2}-a^{2}\cos^{2}\theta
\big[(\hat{L}_{-}-L_{1})^{2}\notag\\&&+(\hat{L}_{-}-L_{2})^{2}\big]\bigg)\bigg]-
\frac{1}{r_{-}^{2}+a^{2}\cos^{2}\theta}\bigg[\cot^{2}\theta L_{1}L_{2}
+a^{2}E^{2}\cos^{2}\theta
+\sqrt{c_{1}(\theta)c_{2}(\theta)}\bigg]\Bigg)\notag\\&&+2m_{1}m_{3}\Bigg(1+\frac{1}{2(\hat{L}_{-}-L_{1})(\hat{L}_{-
}-L_{3})}\bigg[(L_{1}-L_{3})^{2}
+\frac{1}{r_{-}^{2}+a^{2}\cos^{2}\theta}\bigg(\frac{r_{-}^{4}}{a^{2}}
E^{2}\big(L_{1}-L_{3}\big)^{2}\notag\\&&+K_{3}(\hat{L}_{-}-L_{1})^{2}+K_{1}(\hat{L}_{-}-L_{3})^{2}-a^{2}\cos^{2}\theta
\big[(\hat{L}_{-}-L_{1})^{2}+(\hat{L}_{-}-L_{3})^{2}\big]\bigg)\bigg]\notag\\&&\allowdisplaybreaks-
\frac{1}{r_{-}^{2}+a^{2}\cos^{2}\theta}\bigg[\cot^{2}\theta L_{1}L_{3}
+a^{2}E^{2}\cos^{2}\theta
+\sqrt{c_{1}(\theta)c_{3}(\theta)}\bigg]\Bigg)+2m_{2}m_{3}\Bigg(1\notag\allowdisplaybreaks\\&&+\frac{1}{2(\hat{L}_{-}-L_{2})(\hat{L}_{-}
-L_{3})}\bigg[(L_{2}-L_{3})^{2}
+\frac{1}{r_{-}^{2}+a^{2}\cos^{2}\theta}\bigg(\frac{r_{-}^{4}}{a^{2}}
E^{2}\big(L_{2}-L_{3}\big)^{2}+K_{3}(\hat{L}_{-}\notag\allowdisplaybreaks\\&&-L_{2})^{2}+K_{2}(\hat{L}_{-}-L_{3})^{2}-a^{2}\cos^{2}\theta
\big[(\hat{L}_{-}-L_{2})^{2}+(\hat{L}_{-}-L_{3})^{2}\big]\bigg)\bigg]-
\frac{1}{r_{-}^{2}+a^{2}\cos^{2}\theta}\notag\allowdisplaybreaks\\&&\times\bigg[\cot^{2}\theta L_{2}L_{3}
+a^{2}E^{2}\cos^{2}\theta
+\sqrt{c_{2}(\theta)c_{3}(\theta)}\bigg]\Bigg)\Bigg]^{\frac{1}{2}}.
\end{eqnarray}

Let us scrutinize a marginally bound particle $(E=1)$ with the critical angular momentum $\hat{L}_{-}$. The conditions for the admitted region, $b(r)\geq0$ and $c(\theta)\geq0$ grow into
\begin{eqnarray}\label{49}
\cot^{2}\theta\frac{(r_{-}^{2}+a^{2})^{2}}{a^{2}}\leq K\leq \frac{(r+r_{-})^{2}(r-r_{-})}{r-r_{+}}-r^{2}-\frac{r_{-}^{4}}{a^{2}}.
\end{eqnarray}
The inequality (\ref{49}) delivers the upper and lower bounds for the Carter constant $K$. By Eq. (\ref{49}), one can discover the condition for the marginally bound particle with the critical angular momentum to approach the inner horizon of the non-extremal Kerr-MOG black hole
\begin{equation}\label{52}
\Big(\frac{(r+r_{-})^{2}(r-r_{-})}{r-r_{+}}-r^{2}+a^{2}+2r_{-}^{2}\Big)\cos^{2}\theta-\frac{(r+r_{-})^{2}(r-r_{-})}{r-r_{+}}+r^{2}
+\frac{r_{-}^{4}}{a^{2}}\leq0,~~~~\text{for~any}~~~~r\geq r_{-}.
\end{equation}

\begin{figure}[h!]
\centering
\includegraphics[width=10cm, height=10cm]{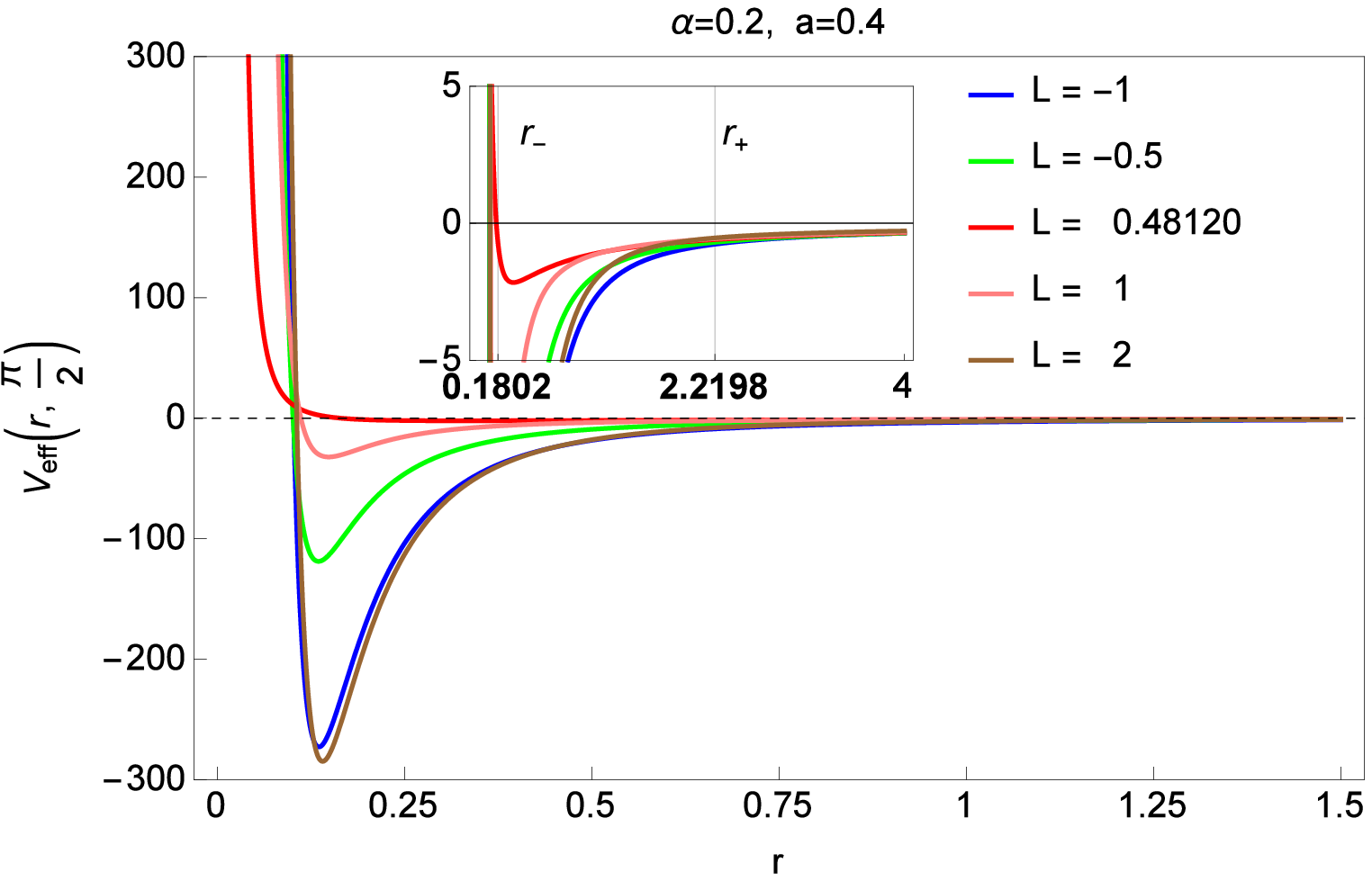}~~~~~~~\\
\includegraphics[width=10cm, height=10cm]{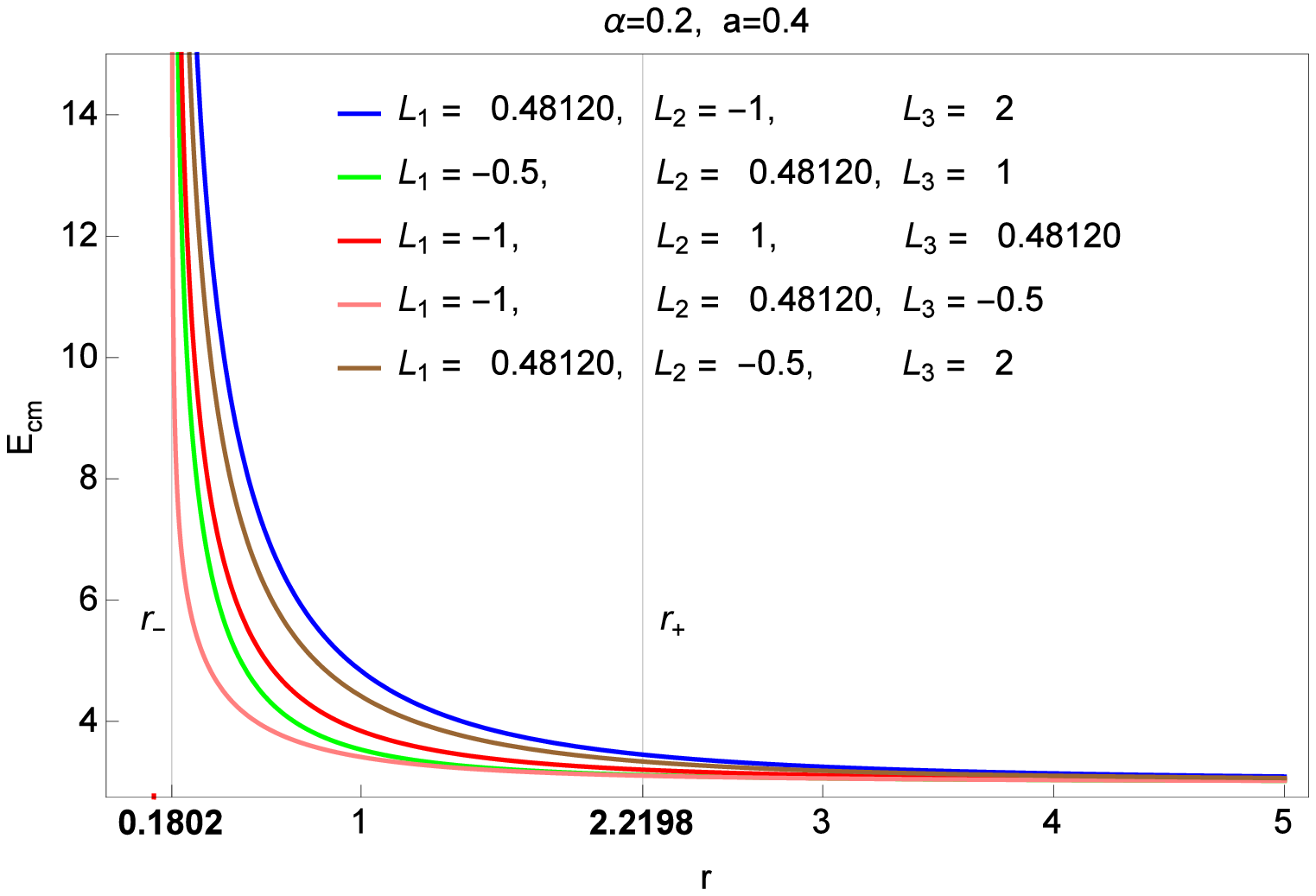}
\caption{The effective potential (top figure) and center of mass energy (bottom figure) for marginally bound particles in the equatorial plane of the non-extremal Kerr-MOG black hole. We set $M=1$, $m_{1}=m_{2}=m_{3}=1$, $a=0.4$, $\alpha=0.2$. Vertical lines identify the place of residence of the inner and outer horizons of the black hole.}\label{f1}
\end{figure}

In the equatorial plane, Eq. (\ref{47}) contributes
\begin{eqnarray}\label{53}
E_{\text{cm}}\Big|_{r\rightarrow r_{-}}&=&\Bigg[m_{1}^{2}+m_{2}^{2}+m_{3}^{2}+2m_{1}m_{2}\Bigg(1+\frac{1}{2(\hat{L}_{-1}-
L_{1})(\hat{L}_{-2}-L_{2})}\bigg(\big[(\hat{L}_{-1}-L_{1})-(\hat{L}_{-2}-L_{2})\big]^{2}\notag\\&&
+\frac{r_{-}^{2}}{(r_{-}^{2}+a^{2})^{2}}
\big(L_{1}\hat{L}_{-2}-L_{2}\hat{L}_{-1}\big)^{2}\bigg)\Bigg)+2m_{1}m_{3}
\Bigg(1+\frac{1}{2(\hat{L}_{-1}-
L_{1})(\hat{L}_{-3}-L_{3})}\bigg(\big[(\hat{L}_{-1}\notag\\&&-L_{1})-(\hat{L}_{-3}-L_{3})\big]^{2}
+\frac{r_{-}^{2}}{(r_{-}^{2}+a^{2})^{2}}
\big(L_{1}\hat{L}_{-3}-L_{3}\hat{L}_{-1}\big)^{2}\bigg)\Bigg)
+2m_{2}m_{3}\Bigg(1\notag\\&&+\frac{1}{2(\hat{L}_{-2}-
L_{2})(\hat{L}_{-3}-L_{3})}\bigg(\big[(\hat{L}_{-2}-L_{2})-(\hat{L}_{-3}-L_{3})\big]^{2}
+\frac{r_{-}^{2}}{(r_{-}^{2}+a^{2})^{2}}
\big(L_{2}\hat{L}_{-3}\notag\\&&-L_{3}\hat{L}_{-2}\big)^{2}\bigg)\Bigg)\Bigg]^{\frac{1}{2}},
\end{eqnarray}
Apparently, the CME is limited for all values of $L_{1}$, $L_{2}$ and $L_{3}$ except when $L_{1}$, $L_{2}$ or $L_{3}$ is approximately equal to the critical angular momentum. For $E_{1}=E_{2}=E_{3}=E$, Eq. (\ref{53}) proceeds the following form
\begin{eqnarray}\label{54}
E_{\text{cm}}\Big|_{r\rightarrow r_{-}}&=&\Bigg[m_{1}^{2}+m_{2}^{2}+m_{3}^{2}+2m_{1}m_{2}\Bigg(1+\frac{1}{2(\hat{L}_{-}-L_{1})(\hat{L}_{-}
-L_{2})}\bigg((L_{1}-L_{2})^{2}+\frac{r_{-}^{2}}{a^{2}}E^{2}\big(L_{1}\notag\\&&-L_{2}\big)^{2}\bigg)\Bigg)
+2m_{1}m_{3}\Bigg(1+\frac{1}{2(\hat{L}_{-}-L_{1})(\hat{L}_{-}-L_{3})}\bigg((L_{1}-L_{3})^{2}
+\frac{r_{-}^{2}}{a^{2}}E^{2}\big(L_{1}-L_{3}\big)^{2}\bigg)\Bigg)\notag\\&&+2m_{2}m_{3}\Bigg(1
+\frac{1}{2(\hat{L}_{-}-L_{2})(\hat{L}_{-}-L_{3})}\bigg((L_{2}-L_{3})^{2}+\frac{r_{-}^{2}}{a^{2}}
E^{2}\big(L_{2}-L_{3}\big)^{2}\bigg)\Bigg)\Bigg]^{\frac{1}{2}}.
\end{eqnarray}
We design the effective potential $V_{\text{eff}}\Big(r,$ {\Large $\frac{\pi}{2}$}$\Big)$ of marginally bound particles in Figure $\ref{f1}$ for $M=1,~a=0.4,~\alpha=0.2$ with different specific angular momenta $L=-1,~-0.5,~\hat{L}_{-},~1,~2$ where $\hat{L}_{-}=0.48120$. Apparently, the effective potential $V_{\text{eff}}\Big(r,$ {\Large $\frac{\pi}{2}$}$\Big)$ is negative for $r\geq r_{-}$, so the particles can approach the inner horizon after overpasses the outer horizon. The subplot expresses the role of $V_{\text{eff}}\Big(r,$ {\Large $\frac{\pi}{2}$}$\Big)$ close to the horizons and identifies the place of residence of the outer and inner horizons. We also design the CME of the collision for $L_{1}=0.48120,~L_{2}=-1,~L_{3}=2$ (Blue Curve), $L_{1}=-0.5,~L_{2}=0.48120,~L_{3}=1$ (Green Curve), $L_{1}=-1,~L_{2}=1,~L_{3}=0.48120$ (Red Curve), $L_{1}=-1,~L_{2}=0.48120,~L_{3}=-0.5$ (Pink Curve), and $L_{1}=0.48120,~L_{2}=-0.5,~L_{3}=2$ (Brown Curve). The CME is limited at the outer horizon and unlimited at the inner horizon $r_{-}=0.1802$.
\subsection{Near-horizon collision of particles around the extremal Kerr-MOG black hole}
Let us deliberate the properties of the CME (\ref{35}) as the particles access the horizon of the extremal Kerr-MOG black hole. In this case $r_{+}=r_{-}=GM$. Using this outcome in Eq. (\ref{39}), we retrieve
\begin{eqnarray}\label{55}
E_{\text{cm}}\Big|_{r\rightarrow GM}&=&\Bigg[m_{1}^{2}+m_{2}^{2}+m_{3}^{2}+2m_{1}m_{2}\Bigg(1+\frac{1}{2(\hat{L}_{1}-L_{1})(\hat{L}_{2}-L_{2})}\bigg[\big[
(\hat{L}_{1}-L_{1})-(\hat{L}_{2}-L_{2})\big]^{2}\notag\\&&+\frac{1}{G^{2}M^{2}+a^{2}\cos^{2}\theta}\bigg(\frac{G^{4}M^{4}}{(G^{2}M^{2}+a^{2})^{2}}
\big(L_{1}\hat{L}_{2}-L_{2}\hat{L}_{1}\big)^{2}+K_{2}(\hat{L}_{1}-L_{1})^{2}+K_{1}(\hat{L}_{2}\notag\\&&-L_{2})^{2}-a^{2}\cos^{2}\theta
\big[(\hat{L}_{1}-L_{1})^{2}+(\hat{L}_{2}-L_{2})^{2}\big]\bigg)\bigg]
-\frac{1}{G^{2}M^{2}+a^{2}\cos^{2}\theta)}\bigg[\cot^{2}\theta L_{1}L_{2}\notag\\&&+\frac{a^{4}\cos^{2}\theta}
{(G^{2}M^{2}+a^{2})^{2}}\hat{L}_{1}\hat{L}_{2}+\sqrt{c_{1}(\theta)c_{2}(\theta)}\bigg]\Bigg)
+2m_{1}m_{3}\Bigg(1+\frac{1}{2(\hat{L}_{1}-L_{1})(\hat{L}_{3}-L_{3})}\notag\\&&\allowdisplaybreaks\times\bigg[\big[
(\hat{L}_{1}-L_{1})-(\hat{L}_{3}-L_{3})\big]^{2}+\frac{1}{G^{2}M^{2}
+a^{2}\cos^{2}\theta}\bigg(\frac{G^{4}M^{4}}{(G^{2}M^{2}+a^{2})^{2}}
\big(L_{1}\hat{L}_{3}-L_{3}\hat{L}_{1}\big)^{2}\notag\\&&+K_{3}(\hat{L}_{1}-L_{1})^{2}+K_{1}(\hat{L}_{3}
-L_{3})^{2}-a^{2}\cos^{2}\theta\big[(\hat{L}_{1}-L_{1})^{2}+(\hat{L}_{3}-L_{3})^{2}\big]\bigg)\bigg]\notag\\&&-
\frac{1}{G^{2}M^{2}+a^{2}\cos^{2}\theta)}\bigg[\cot^{2}\theta L_{1}L_{3}+\frac{a^{4}\cos^{2}\theta}
{(G^{2}M^{2}+a^{2})^{2}}\hat{L}_{1}\hat{L}_{3}+\sqrt{c_{1}(\theta)c_{3}(\theta)}\bigg]\Bigg)
\notag\\&&+2m_{2}m_{3}\Bigg(1+\frac{1}{2(\hat{L}_{2}-L_{2})(\hat{L}_{3}-L_{3})}\bigg[\big[
(\hat{L}_{2}-L_{2})-(\hat{L}_{3}-L_{3})\big]^{2}\notag\\&&+\frac{1}{G^{2}M^{2}
+a^{2}\cos^{2}\theta}\bigg(\frac{G^{4}M^{4}}{(G^{2}M^{2}+a^{2})^{2}}
\big(L_{2}\hat{L}_{3}-L_{3}\hat{L}_{2}\big)^{2}+K_{3}(\hat{L}_{2}-L_{2})^{2}+K_{2}(\hat{L}_{3}
\notag\\&&-L_{3})^{2}-a^{2}\cos^{2}\theta
\big[(\hat{L}_{2}-L_{2})^{2}+(\hat{L}_{3}-L_{3})^{2}\big]\bigg)\bigg]-
\frac{1}{G^{2}M^{2}+a^{2}\cos^{2}\theta)}\bigg[\cot^{2}\theta L_{2}L_{3}\notag\\&&+\frac{a^{4}\cos^{2}\theta}
{(G^{2}M^{2}+a^{2})^{2}}\hat{L}_{2}\hat{L}_{3}+\sqrt{c_{2}(\theta)c_{3}(\theta)}\bigg]\Bigg)\Bigg]^{\frac{1}{2}}.
\end{eqnarray}
where $\hat{L}_{i}=G_{N}GM^{2}(2+\alpha)\frac{E_{i}}{a}$ is the critical angular momentum at the horizon for the $i$th particle. The essential condition for achieving an arbitrarily huge CME is $L_{i}=\hat{L}_{i}$ for either of the three particles. For $E_{1}=E_{2}=E_{3}=E$, we attain the identical critical angular momentum i.e., $\hat{L}_{i}=G_{N}GM^{2}(2+\alpha)\frac{E}{a}$ , and Eq. (\ref{55}) gives
\begin{eqnarray}\label{56}
E_{\text{cm}}\Big|_{r\rightarrow GM}&=&\Bigg[m_{1}^{2}+m_{2}^{2}+m_{3}^{2}+2m_{1}m_{2}\Bigg(1+\frac{1}{2(\hat{L}-L_{1})(\hat{L}-L_{2})}\bigg[(L_{1}
-L_{2})^{2}+\frac{1}{G^{2}M^{2}+a^{2}\cos^{2}\theta}\notag\\&&\times\bigg(\frac{G^{4}M^{4}}{a^{2}}
E^{2}\big(L_{1}-L_{2}\big)^{2}+K_{2}(\hat{L}-L_{1})^{2}+K_{1}(\hat{L}-L_{2})^{2}-a^{2}\cos^{2}\theta
\big[(\hat{L}-L_{1})^{2}+(\hat{L}\notag\\&&-L_{2})^{2}\big]\bigg)\bigg]-
\frac{1}{G^{2}M^{2}+a^{2}\cos^{2}\theta}\bigg[\cot^{2}\theta L_{1}L_{2}+
a^{2}\cot^{2}\theta E^{2}
+\theta\sqrt{c_{1}(\theta)c_{2}(\theta)}\bigg]\Bigg)\notag\\&&+2m_{1}m_{3}\Bigg(1+\frac{1}{2(\hat{L}-L_{1})(\hat{L}
-L_{3})}\bigg[(L_{1}-L_{3})^{2}+\frac{1}{G^{2}M^{2}+a^{2}\cos^{2}\theta}\bigg(\frac{G^{4}M^{4}}{a^{2}}
E^{2}\big(L_{1}\notag\\&&-L_{3}\big)^{2}+K_{3}(\hat{L}-L_{1})^{2}+K_{1}(\hat{L}-L_{3})^{2}-a^{2}\cos^{2}\theta
\big[(\hat{L}-L_{1})^{2}+(\hat{L}-L_{3})^{2}\big]\bigg)\bigg]\notag\\&&-\frac{1}{G^{2}M^{2}
+a^{2}\cos^{2}\theta}\bigg[\cot^{2}\theta L_{1}L_{3}+a^{2}\cot^{2}\theta E^{2}
+\theta\sqrt{c_{1}(\theta)c_{3}(\theta)}\bigg]\Bigg)+2m_{2}m_{3}\Bigg(1\notag\\&&+\frac{1}{2(\hat{L}-L_{2})(\hat{L}
-L_{3})}\bigg[(L_{2}-L_{3})^{2}+\frac{1}{G^{2}M^{2}+a^{2}\cos^{2}\theta}\bigg(\frac{G^{4}M^{4}}{a^{2}}
E^{2}\big(L_{2}-L_{3}\big)^{2}\notag\\&&+K_{3}(\hat{L}-L_{2})^{2}+K_{2}(\hat{L}-L_{3})^{2}-a^{2}\cos^{2}\theta
\big[(\hat{L}-L_{2})^{2}+(\hat{L}-L_{3})^{2}\big]\bigg)\bigg]\notag\\&&-\frac{1}{G^{2}M^{2}
+a^{2}\cos^{2}\theta}\bigg[\cot^{2}\theta L_{2}L_{3}+a^{2}\cot^{2}\theta E^{2}
+\theta\sqrt{c_{2}(\theta)c_{3}(\theta)}\bigg]\Bigg)\Bigg]^{\frac{1}{2}}.
\end{eqnarray}
Let us assume a marginally bound particle $(E=1)$ with the critical angular momentum $\hat{L}$. The inequality (\ref{41}) gives
\begin{equation}\label{57}
 G_{N}GM^{2}(2+\alpha)^{2}\cot^{2}\theta \leq K\leq GM(2r-\alpha GM).
\end{equation}
Thus for the marginally bound particle with the critical angular momentum to approach the horizon of the extremal Kerr-MOG black hole, the following circumstances must be fulfilled
\begin{equation}\label{60}
\big(G_{N}M(2+\alpha)^{2}+2r-\alpha GM\big)\cos^{2}\theta-2r+\alpha GM\leq0,~~~~~~~~~~~~\text{for~any}~~~~~~r\geq GM.
\end{equation}

In addition, if the collision arises in the equatorial plane, the CME (\ref{55}) at the horizon of the extremal Kerr-MOG black hole becomes
\begin{eqnarray}\label{61}
E_{\text{cm}}\Big|_{r\rightarrow GM}&=&\Bigg[m_{1}^{2}+m_{2}^{2}+m_{3}^{2}+2m_{1}m_{2}\Bigg(1+\frac{1}{2(\hat{L}_{1}-L_{1})(\hat{L}_{2}-L_{2})}\bigg(\big[
(\hat{L}_{1}-L_{1})-(\hat{L}_{2}-L_{2})\big]^{2}\notag\allowdisplaybreaks\\&&+\frac{G^{2}M^{2}}{(G^{2}M^{2}+a^{2})^{2}}
\big(L_{1}\hat{L}_{2}-L_{2}\hat{L}_{1}\big)^{2}\bigg)
\Bigg)+2m_{1}m_{3}\Bigg(1+\frac{1}{2(\hat{L}_{1}-L_{1})(\hat{L}_{3}-L_{3})}\notag\\&&\allowdisplaybreaks\times\bigg(\big[
(\hat{L}_{1}-L_{1})-(\hat{L}_{3}-L_{3})\big]^{2}+\frac{G^{2}M^{2}}{(G^{2}M^{2}+a^{2})^{2}}
\big(L_{1}\hat{L}_{3}-L_{3}\hat{L}_{1}\big)^{2}\bigg)\Bigg)
+2m_{2}m_{3}\Bigg(1\notag\allowdisplaybreaks\\&&+\frac{1}{2(\hat{L}_{2}-L_{2})(\hat{L}_{3}-L_{3})}\bigg(\big[
(\hat{L}_{2}-L_{2})-(\hat{L}_{3}-L_{3})\big]^{2}+\frac{G^{2}M^{2}}{(G^{2}M^{2}+a^{2})^{2}}
\big(L_{2}\hat{L}_{3}\notag\allowdisplaybreaks\\&&-L_{3}\hat{L}_{2}\big)^{2}\bigg)\Bigg)\Bigg]^{\frac{1}{2}}.
\end{eqnarray}
which is undeniably finite for all values of $L_{1}$, $L_{2}$ and $L_{3}$ except when $L_{1}$, $L_{2}$ or $L_{3}$ approaches the critical angular momentum, for which the CME is arbitrarily huge. When the specific energy of all the three particles are absolutely identical, then (\ref{61}) diminishes to
\begin{eqnarray}\label{62}
E_{\text{cm}}\Big|_{r\rightarrow GM}&=&\Bigg[m_{1}^{2}+m_{2}^{2}+m_{3}^{2}+2m_{1}m_{2}\Bigg(1+\frac{1}{2(\hat{L}-L_{1})(\hat{L}-L_{2})}\bigg((L_{1}
-L_{2})^{2}+\frac{G^{2}M^{2}}{a^{2}}
E^{2}\big(L_{1}\notag\\&&-L_{2}\big)^{2}\bigg)\Bigg)+2m_{1}m_{3}\Bigg(1+\frac{1}{2(\hat{L}-L_{1})(\hat{L}
-L_{3})}\bigg((L_{1}-L_{3})^{2}+\frac{G^{2}M^{2}}{a^{2}}
E^{2}\big(L_{1}\notag\\&&-L_{3}\big)^{2}\bigg)\Bigg)+2m_{2}m_{3}\Bigg(1+\frac{1}{2(\hat{L}-L_{2})(\hat{L}
-L_{3})}\bigg((L_{2}-L_{3})^{2}+\frac{G^{2}M^{2}}{a^{2}}
E^{2}\big(L_{2}\notag\\&&-L_{3}\big)^{2}\bigg)\Bigg)\Bigg]^{\frac{1}{2}}.
\end{eqnarray}
We design the effective potential $V_{\text{eff}}\Big(r,$ {\Large $\frac{\pi}{2}$}$\Big)$ of marginally bound particles in Figure $\ref{f1}$ for $M=1,~a=1.41421,~\alpha=1$ with different specific angular momenta $L=-1,~-0.5,~1,~2.5,~4.24264$ where $4.24264$ is the critical angular momentum. Apparently, the effective potential $V_{\text{eff}}\Big(r,$ {\Large $\frac{\pi}{2}$}$\Big)$ is negative for $r\geq 2$, so the particles can approach the horizon. The subplot expresses the role of $V_{\text{eff}}\Big(r,$ {\Large $\frac{\pi}{2}$}$\Big)$ close to the horizon and identifies the place of residence of the horizon. We also design the CME of the collision for $L_{1}=4.24264,~L_{2}=-1,~L_{3}=2.5$ (Blue Curve), $L_{1}=-0.5,~L_{2}=4.24264,~L_{3}=2.5$ (Green Curve), $L_{1}=-1,~L_{2}=1,~L_{3}=4.24264$ (Red Curve), $L_{1}=-1,~L_{2}=4.24264,~L_{3}=-0.5$ (Pink Curve), and $L_{1}=4.24264,~L_{2}=1,~L_{3}=2.5$ (Brown Curve). The CME is unlimited at the horizon $r_{+}=r_{-}=2$.
\begin{figure}[h!]
\includegraphics[width=10cm, height=10cm]{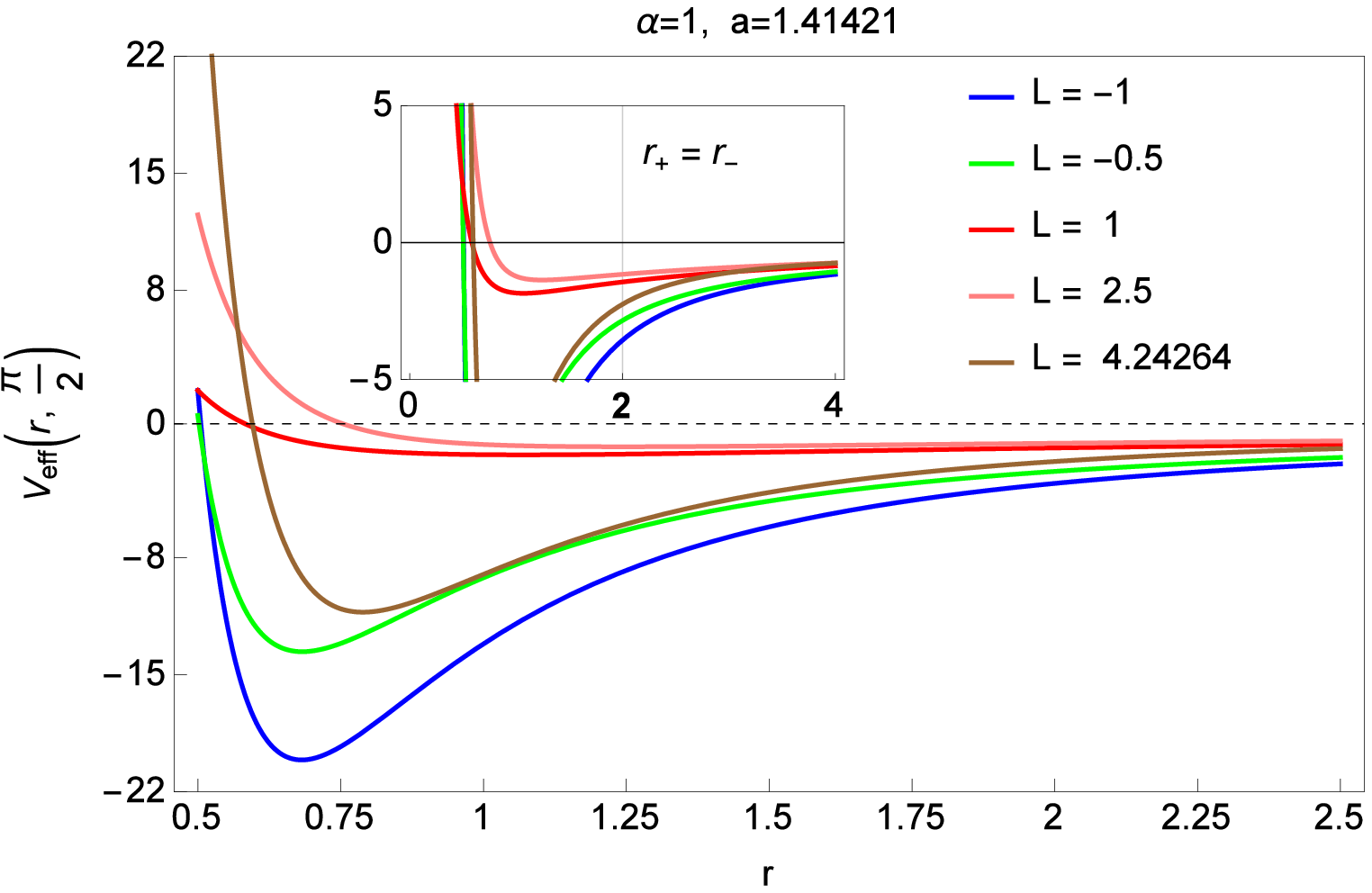}~~~~~~~\\
\includegraphics[width=10cm, height=10cm]{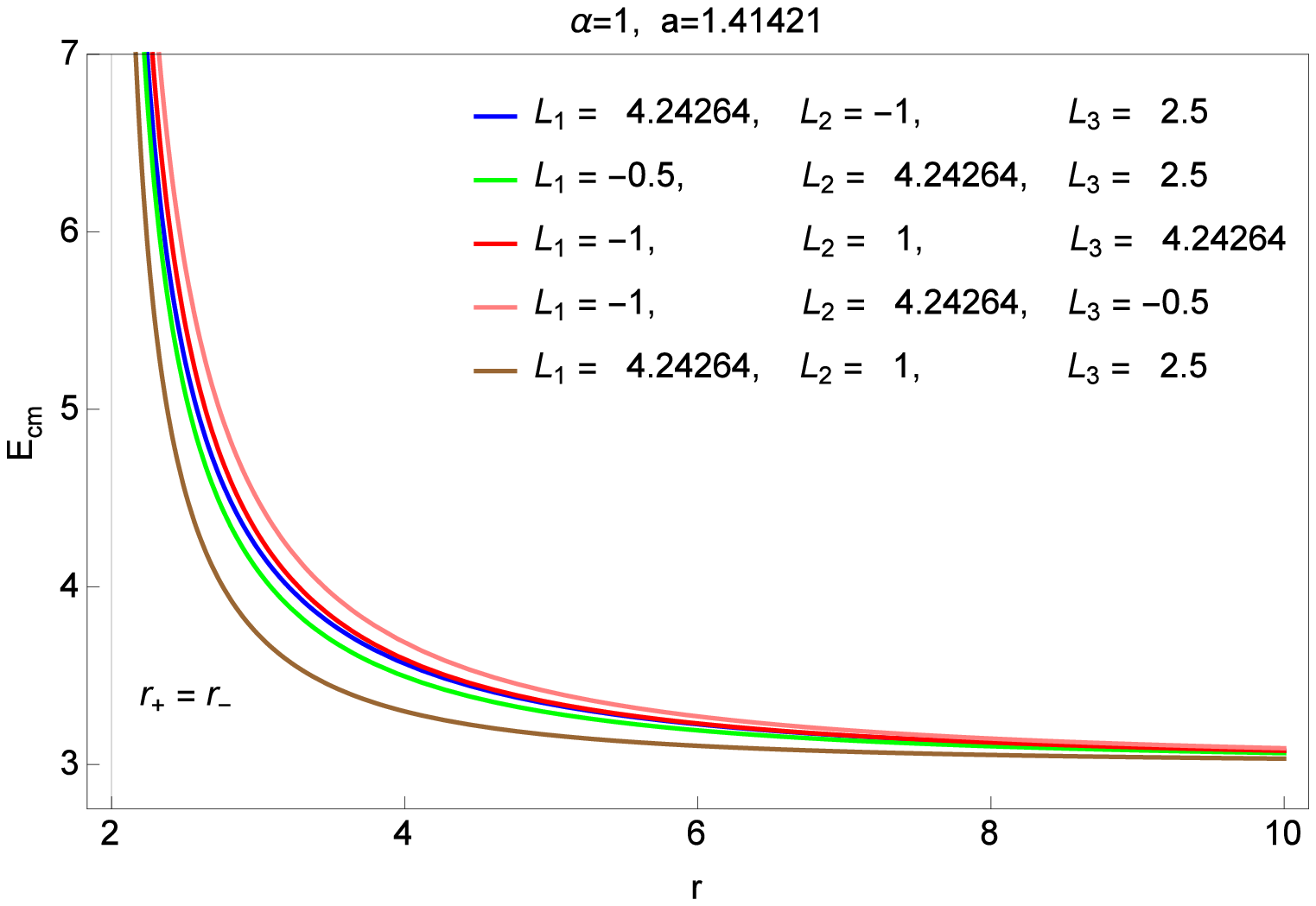}
\caption{The effective potential (top figure) and center of mass energy (bottom figure) for marginally bound particles in the equatorial plane of the extremal Kerr-MOG black hole. We set $M=1$, $m_{1}=m_{2}=m_{3}=1$, $a=1.41421$, $\alpha=1$. Vertical lines identify the place of residence of the horizon of the black hole.}\label{f2}
\end{figure}
\newpage
\section{Conclusion}
In this paper, we have reviewed the CME of the collision for three neutral particles with different
rest masses falling freely from rest at infinity in the vicinity of a Kerr-MOG black hole. In addition, we have deliberated the CME when the collision takes place near the horizon(s) of an extremal and non-extremal Kerr-MOG black hole. We have discovered that an arbitrarily huge CME is attainable with following conditions: (1) the collision arises at the horizon(s) of an extremal and non-extremal Kerr-MOG black hole, (2) the spin parameter $a\neq0$, and (3) one of the colliding particles has critical angular momentum. We explored the upper and lower bounds of the Carter constant $K$ for a marginally bound particle with the critical angular momentum in an extremal and non-extremal Kerr-MOG black hole. In the equatorial plane, we revealed that there exist intervals for the specific angular momentum $L$ correspond to the spin parameter $a$ and gravitational field strength $\alpha$ for which not only three marginally bound particles approach the horizons of the non-extremal Kerr-MOG black hole but also the collision of these particles takes place at the horizons.


\begin{thebibliography}{99}
\bibitem{1} M. Ba\~{n}ados, J. Silk, S. M. West, Phys. Rev. Lett. {\bfseries{103}}, 111102 (2009).
\bibitem{2} A. A. Grib, Yu. V. Pavlov, Astropart. Phys. {\bfseries{34}}, 581 (2011).
\bibitem{3} A. A. Grib, Yu. V. Pavlov, JETP Lett. {\bfseries{92}}, 125 (2010).
\bibitem{4} A. A. Grib, Yu. V. Pavlov, Grav. Cosmol. {\bfseries{17}}, 42 (2011).
\bibitem{5} E. Berti, V. Cardoso, L. Gualtieri, F. Pretorius, U. Sperhake, Phys. Rev. Lett. {\bfseries{103}}, 239001 (2009).
\bibitem{6} T. Jacobson, T. P. Sotiriou, Phys. Rev. Lett. {\bfseries{104}}, 021101 (2010).
\bibitem{7} K. Lake, Phys. Rev. Lett. {\bfseries{104}}, 211102 (2010).
\bibitem{8} T. Harada, M. Kimura, Phys. Rev. {\bfseries{D}} {\bfseries{83}}, 024002 (2011).
\bibitem{9} T. Harada, M. Kimura, Class. Quantum Grav. {\bfseries{31}}, 243001 (2014).
\bibitem{10} A. Galajinsky, Phys. Rev. {\bfseries{D}} {\bfseries{88}}, 027505 (2013).
\bibitem{11} T. Harada, M. Kimura, Phys. Rev. {\bfseries{D}} {\bfseries{83}}, 084041 (2011).
\bibitem{12} O. B. Zaslavskii, Phys. Rev. {\bfseries{D}} {\bfseries{82}}, 083004 (2010).
\bibitem{13} C. Liu, S. Chen, J. Jing, Chin. Phys. Lett. {\bfseries{30}}, 100401 (2013).
\bibitem{14} S. W. Wei, Y. X. Liu, H. Guo, C. E. Fu, Phys. Rev. {\bfseries{D}} {\bfseries{82}}, 103005 (2010).
\bibitem{15} V. P. Frolov, Phys. Rev. {\bfseries{D}} {\bfseries{85}}, 024020 (2012).
\bibitem{16} C. Liu, S. Chen, C. Ding, J. Jing, Phys. Lett. {\bfseries{B}} {\bfseries{701}}, 285 (2011).
\bibitem{17} C. Chakraborty, Eur. Phys. J. {\bfseries{C}} {\bfseries{74}}, 2759 (2014).
\bibitem{18} C. Chakraborty, P. Majumdar, Class. Quantum Grav. {\bfseries{31}}, 075006 (2014).
\bibitem{19} C. Chakraborty, Eur. Phys. J. {\bfseries{C}} {\bfseries{75}}, 572 (2015).
\bibitem{20} I. Hussain, Mod. Phys. Lett. {\bfseries{A}} {\bfseries{27}}, 1250017 (2012).
\bibitem{21} A. Zakria, M. Jamil, JHEP {\bfseries{5}}, 147 (2015).
\bibitem{22} I. Hussain, J. Phys.: Conf. Ser. {\bfseries{354}}, 012007 (2012).
\bibitem{23} S. W. Wei, Y. X. Liu, H. T. Li, F. W. Chen, JHEP {\bfseries{12}}, 066 (2010).
\bibitem{24} M. Jamil, S. Hussain, B. Majeed, Eur. Phys. J. {\bfseries{C}} {\bfseries{75}}, 24 (2015).
\bibitem{25} A. Tursunov, M. Kolos, A. Abdujabbarov, B. Ahmedov, and Z. Stuchlik, Phys. Rev. {\bfseries{D}} {\bfseries{88}}, 124001 (2013).
\bibitem{26} I. Hussain, M. Jamil, B. Majeed, Int. J. Theor. Phys. {\bfseries{54}}, 1567 (2015).
\bibitem{27} A. A. Abdujabbarov, A.A. Tursunov, B.J. Ahmedov, A. Kuvatov, Astrophys. Space Sci. {\bfseries{343}}, 173 (2013).
\bibitem{28} O. B. Zaslavskii, JETP Lett. {\bfseries{92}}, 571 (2010).
\bibitem{29} N. Haider, Open J. Mod. Phys. {\bfseries{1}}, 34 (2014).
\bibitem{30} O. B. Zaslavskii, Class. Quantum Grav. {\bfseries{28}}, 105010 (2011).
\bibitem{31} M. Sharif, N. Haider, Astrophys. Space Sci. {\bfseries{346}}, 111 (2013).
\bibitem{32} I. Hussain, Mod. Phys. Lett. {\bfseries{A}} {\bfseries{27}}, 1250068 (2012).
\bibitem{33} S. Hussain, M. Jamil, Phys. Rev. {\bfseries{D}} {\bfseries{92}}, 043008 (2015).
\bibitem{34} U. Debnath, arXiv:1508.02385.
\bibitem{35} M. Halilsoy, A. Ovgun, Cana. J. Phys. ja (2015).
\bibitem{36} B. Pourhassan, U. Debnath, arXiv:1506.03443.
\bibitem{37} A. Zakria,  Q. Satti,	arXiv:1807.01621.
\bibitem{38} J. W. Moffat, Eur. Phys. J. {\bfseries{C}} {\bfseries{75}}, 175 (2015).
\bibitem{39} M. Sharif, M. Shahzadi, Eur. Phys. J. {\bfseries{C}} {\bfseries{77}}, 363 (2017).
\end{thebibliography}
\end{document}